\documentclass[aps,pra,twocolumn,amsmath,amssymb,nofootinbib,showpacs,superscriptaddress]{revtex4-1}
\usepackage[english]{babel}
\usepackage{latexsym}
\usepackage{graphics}
\usepackage{graphicx}
\usepackage{epsfig}
\usepackage{color}
\usepackage{bm}
\usepackage{amsmath}
\usepackage{amssymb}
\usepackage{amsthm}
\usepackage{dcolumn}
\usepackage{bbm}
\usepackage{float}
\usepackage{hyperref}
\usepackage{color}
\usepackage{epstopdf}
\usepackage{cleveref}
\usepackage[svgnames]{xcolor}
\usepackage{physics}
\usepackage[sort&compress]{natbib}
\usepackage{bbold}
\usepackage{enumerate}

\hypersetup{hidelinks,colorlinks=true,allcolors=DarkBlue}

\newcommand{\BLUE}[1]{#1  }

\newcommand{\Neg}{{\cal N}}
\newcommand{\NegTot}{{\cal N}_{\rm tot}}
\newcommand{\Abs}[1]{\left| #1 \right|}
\newcommand{\Norm}[1]{\left\lVert #1 \right\rVert}
\newcommand{\Hil}{\mathcal{H}}
\newcommand{\THil}{\mathfrak{T}(\Hil)}
\newcommand{\BHil}{\mathfrak{B}(\Hil)}
\newcommand{\lOne}{\ell_1}
\newcommand{\lInf}{\ell_\infty}
\newcommand{\Id}{\mathrm{id}}
\newcommand{\Def}{:=}
\newcommand{\Idop}{\hat{\mathbb{1}}}
\newcommand{\POVM}{{\cal F}}
\newcommand{\POVMEff}{\hat{F}}
\newcommand{\Chan}{\Phi}
\newcommand{\State}{\hat{\rho}}
\newcommand{\ProbVecP}{p}

\newcommand{\ProbVecV}{v}
\newcommand{\MeasVecV}{v}
\newcommand{\QSMatr}{S}
\newcommand{\QSMatrM}{V}
\newcommand{\EMap}{{\bf E}}
\newcommand{\eMap}{{\bf e}}
\newcommand{\eOp}{\hat{e}}
\newcommand{\EOp}{\hat{E}}
\newcommand{\Traj}{{\boldsymbol x}}
\newcommand{\TrajSet}{{\bf X}}
\newcommand{\OutProb}{q_{\rm out}}
\newcommand{\OutProbEst}{\OutProb^{\rm est}}
\newcommand{\Stoch}{{\rm st}}
\newcommand{\Circ}{{\cal C}}

\begin{document}
\preprint{APS/123-QED}

\title{Minimizing the negativity of quantum circuits in \texorpdfstring{\\}{} overcomplete quasiprobability representations}

\author{Denis~A. Kulikov}
\affiliation{Russian Quantum Center, Skolkovo, Moscow 121205, Russia}
\affiliation{Moscow Institute of Physics and Technology, Dolgoprudny 141700, Russia}
\affiliation{National University of Science and Technology ``MISIS'', Moscow 119049, Russia}

\author{Vsevolod~I. Yashin}
\affiliation{Steklov Mathematical Institute of Russian Academy of Sciences, Moscow 119991, Russia}
\affiliation{Russian Quantum Center, Skolkovo, Moscow 121205, Russia}
\affiliation{National University of Science and Technology ``MISIS'', Moscow 119049, Russia}

\author{Aleksey~K. Fedorov}
\affiliation{Russian Quantum Center, Skolkovo, Moscow 121205, Russia}
\affiliation{National University of Science and Technology ``MISIS'', Moscow 119049, Russia}

\author{Evgeniy~O. Kiktenko}
\affiliation{Russian Quantum Center, Skolkovo, Moscow 121205, Russia}
\affiliation{National University of Science and Technology ``MISIS'', Moscow 119049, Russia}
\affiliation{Steklov Mathematical Institute of Russian Academy of Sciences, Moscow 119991, Russia}

\begin{abstract}
The problem of simulatability of quantum processes using classical resources plays a cornerstone role for quantum computing. 
Quantum circuits can be simulated classically, e.g., using Monte Carlo sampling techniques applied to quasiprobability representations of circuits' basic elements, i.e., states, gates, and measurements.
The effectiveness of the simulation is determined by the amount of the negativity in the representation of these basic elements.
Here we develop an approach for minimizing the total negativity of a given quantum circuit with respect to quasiprobability representations, that are overcomplete, 
i.e., are such that the dimensionality of corresponding quasistochastic vectors and matrices is larger than the squared dimension of quantum states.
Our approach includes both optimization over equivalent quasistochastic vectors and matrices, which appear due to the overcompleteness, and optimization over overcomplete frames.
We demonstrate the performance of the developed approach on some illustrative cases, and show its significant advantage compared to the standard overcomplete quasistochastic representations.
We also study the negativity minimization of noisy brick-wall random circuits via a combination of increasing frame dimension and applying gate merging technique. 
We demonstrate that the former approach appears to be more efficient in the case of a strong decoherence.
\end{abstract}

\maketitle

\section{Introduction} \label{sec:introduction}

The problem of classical simulation of quantum dynamics is crucial for the understanding of a potential computational speedup (or lack there of) of quantum computing devices compared to systems based on classical principles.
Modern algorithms for classical simulation of quantum circuits predict the exponential growth of computational cost depending on the amount of certain quantum resource,
which can be formalized via the number of T gates or magic states~\cite{Bravyi_2016, Braviy_2019, Howard_2017}, contextuality~\cite{Howard_2014, Delfosse_2015, Delfosse_2017}, 
or negativity~\cite{Spekkens_2008, Veitch_2012, Veitch_2014,Delfosse_2017}.
The latter one seems to be the most general and natural resource describing the classical complexity of quantum dynamics, since it applies to any quasiprobability representation of quantum mechanics~\cite{Pashayan_2015}.
Being, at first sight, a fundamental problem, the question of a relation between available quantum resources and potential advantages of quantum systems in various tasks such as solving computational problems~\cite{Preskill2018},
has become more practical since currently available quantum computing devices are significantly affected by noise~\cite{Preskill2018,Aspuru-Guzik2021,Babbush2021-4}. 
Currently available noisy intermediate-scale quantum (NISQ) devices compete with modern computers based on classical principles~\cite{Martinis2019,Pan2020,Pan2021-4,Pan2021-5},
and the role of quantum effects in these experiments is of significant interest for understanding the origin of quantum advantage. 

Quasiprobability representations of quantum systems provide a general scheme for representing quantum states as quasidistributions and channels as quasistochastic matrices --- 
analogs of standard probability distributions and stochastic matrices with omitted constraint on nonnegativity of the elements~\cite{Ferrie_2008, Ferrie_2009, Ferrie_2011}.
The well-known examples include the Wigner function~\cite{Wigner_1932}, discrete Wigner functions~\cite{Gross_2006, Delfosse_2015},
tomographic representations~\cite{DAriano_2001, DAriano_2004, Amosov_2018, andreev2020quantizer, man2021probability}, 
in particular symmetric, informationally complete positive operator-valued measure (SIC-POVM) representation\cite{Renes_2004, filippov2010symmetric, van2017quantum, appleby2017introducing, Kiktenko_2020, DeBrota_2020}.
Using a quasiprobability representation, one can estimate a probability of a given measurement outcome for a given quantum circuit via Monte Carlo procedure~\cite{Pashayan_2015}.
Notably, the rate of convergence depends on the negativity of occurring quasidistributions and quasistochastic matrices.
It has been shown that negativity is necessary for describing general quantum operations~\cite{Ferrie_2008}.
However, recently quasiprobability representations with zero negativity on Clifford+Magic circuits~\cite{Zurel_2020,Okay_2021,Zurel_2021,Raussendorf_2020} have been found: any quantum state is represented positively, but non-Pauli measurements can have negativity. 
In such representations the dimensionality grows superexponentially with the number of qubits.
Within this context, an important question is to find a quasiprobability representations, which is optimal in the sense of minimizing the negativity. 
A number of recent studies have been dedicated to finding such optimal representations, which would minimize the occurring negativities for some circuits~\cite{Yashin_2020,Koukoulekidis_2022,Luchnikov_2019}.

In the current paper, we develop a generalized framework of quasiprobability representations aimed at representing a given element of a circuit (state, channel, or measurement) in terms of quasiprobabilities with the minimal negativity.
The main feature of the developed approach is its possibility to operate in the overcomplete case, where quasistochastic vectors and matrices are of arbitrary dimensionality $M>d^2$, and $d$ is the dimension of the original Hilbert state space.
We generalize the known frame-based approach~\cite{Ferrie_2008, Ferrie_2009, Ferrie_2011, Pashayan_2015, DAriano_2004} and specify the overcomplete representation by synthesis map only, and obtain an additional room for improving negativity:
The same object (quantum state or channel) can have a number of equivalent representations, and we are able to choose the one with the minimal negativity.
We employ this framework to the problem of minimizing the total negativity of a particular circuit with a given number of particular elements.
In the result, we obtain an algorithm that combines optimization over synthesis maps of specified dimensionality $M$ and optimization of circuit element representation with respect to the fixed synthesis map.
We focus on factorized representations that provide a straightforward way to operate with multi-particle (e.g., multiqubit) systems.
We apply the resulting negativity minimization algorithm to illustrative cases of (noisy) Clifford+T gates, and (noisy) variational circuits.
We demonstrate the decrease of the negativity with the growth of $M$ and decoherence strength, and also show an advantage of our optimization results compared to standard overcomplete representations that have been considered in the literature.
We also consider a combination of increasing frame dimension with gate merging technique, which has been developed in Ref.~\cite{Koukoulekidis_2022}, for minimizing negativity of  noisy brick-wall random circuits.

The paper is organized as follows.
In Sec.~\ref{sec:representations}, we discuss the standard frame-based quasiprobability representations and recall the importance of negativity to the classical simulation problem.
In Sec.~\ref{sec:general_representations}, we introduce generalized quasiprobability representations and show how to find the representation of elements (states, channels, measurements) with minimal negativity.
In Sec.~\ref{sec:optimization}, we develop an algorithm for finding the optimal quasiprobability representation for a given quantum circuit.
Then in Sec.~\ref{sec:performance} we test the algorithm on some simple circuits.
In Sec.~\ref{sec:random_circs}, we combine our approach with the gate merging technique for minimizing negativity of noisy random brick-wall circuits.
We summarize main results and conclude in Sec.~\ref{sec:conclusion}.

\section{Quasiprobability representations and negativity} \label{sec:representations}

\subsection{Basic definitions and notations} \label{subsubsec:notation}

We start by introducing basic notations and definitions that are used throughout the paper.
Let $\Hil$ be a Hilbert space of some finite dimension $d$.
Let $\THil$ be a space of trace-class operators on $\Hil$ equipped with the trace norm $\Norm{\,\cdot\,}_1 \Def \Tr\Abs{\,\cdot\,}$.
We call self-adjoint and unit-trace operator $\State\in\THil$ ($\State=\State^\dagger$, $\Tr[\State]=1$) a \emph{quasistate}.
Moreover, if $\State\geq 0$ then $\State$ is called a \emph{quantum state} or, simply, a \emph{state}.

Let $\BHil$ be a space of operators on $\Hil$ equipped with the operator norm defined for $\POVMEff \in \BHil$ as $\| \POVMEff \|_\infty \Def \sup
\{ \| \POVMEff \psi \| : \Norm{\psi} \leq 1,\, \psi\in \Hil \}$.
We call a self-adjoint operator $\POVMEff \in\BHil$ an \emph{observable} over states from $\Hil$.
If certain $\POVMEff \in\BHil$ also satisfies $0\leq \POVMEff \leq \Idop$, where $\Idop$ is the identity operator, then we call $\POVMEff$ an \emph{effect}.
A finite set of effects $\{\POVMEff_k\}_{k=1}^K$ whose elements sum to the identity operator ($\sum_k \POVMEff_k=\Idop$) is called a positive operator-valued measurement (POVM).
We denote a linear dimension of $\THil$ and $\BHil$ as $D\Def d^2$.
The spaces of states $\THil$ and observables $\BHil$ are in the duality given by the pairing (nondegenerate bilinear map)
\begin{equation} \label{eq:duality_quant}
	(\State, \POVMEff) \mapsto {\Tr} (\State\,\POVMEff) ~\text{for}~\State \in\THil~\text{and}~ \POVMEff\in\BHil.
\end{equation}
Recall that for a state $\State$ and an effect $\POVMEff$, this pairing provides a probability of obtaining the corresponding measurement outcome.
We note that although formally in the finite-dimensional case $\THil$ and $\BHil$ are the same sets, their elements describe different physical procedures of state preparation and measurement correspondingly.

To describe an evolution of states, we employ the concept of \emph{quantum channels} that are completely positive trace-preserving (CPTP) maps $\THil \rightarrow \THil$, 
which we usually denote as $\Chan$ (we consider only the case where the dimensions of input and output state spaces are equal).

Let us then introduce ``classical'' real valued counterparts for $\THil$ and $\BHil$.
We denote $\lOne^M$ the space of $M$-length real-valued tuples equipped with the $1$-norm $\Norm{\ProbVecP}_1 = \sum_{m=1}^M \Abs{\ProbVecP_m}$, where $\ProbVecP_m$ with $m=1,\ldots,M$ denotes elements of $\ProbVecP\in \lOne^M$
For the sake of readability we may omit $M$ and write $\lOne$.
For our purposes, it is convenient to treat $\ProbVecP \in \lOne$ as a column vector.
We call $\ProbVecP \in \lOne$ satisfying  normalization condition $\sum_m p_m=1$ a \emph{quasiprobability distribution} or a \emph{quasidistribution}.
In the special case, where all elements of $\ProbVecP$ are also non-negative ($\ProbVecP_m\geq 0$), we call $\ProbVecP$ a \emph{probability distribution} or a \emph{distribution}.

We denote $\lInf^M$ (or $\lInf$) a space of $M$-length real-valued tuples with the $\infty$-norm $\Norm{\MeasVecV}_\infty \Def \max_{m=1}^M \Abs{\MeasVecV_m}$ for $\MeasVecV=(\MeasVecV_1,\ldots,\MeasVecV_M)\in\lInf^M$.
An element $\MeasVecV\in\lInf$ can be considered as a classical observable.
Note that if all entries are non-negative ($\MeasVecV_m \in [0,1]$), then $\MeasVecV$ may be thought of as a \emph{fuzzy set}~\cite{Zadeh_1965}.
The spaces $\lOne^M$ and $\lInf^M$ are in the duality given by the nondegenerate bilinear map
\begin{equation} \label{eq:duality_class}
    (\ProbVecP,\MeasVecV) \mapsto \sum_m \MeasVecV_m \ProbVecP_m = \MeasVecV\,\ProbVecP\quad \text{for}~ \ProbVecP \in\lOne^M~\text{and}~ \MeasVecV \in\lInf^M,
\end{equation}
where we treat $\MeasVecV$ as a row vector.
One can see that \eqref{eq:duality_class} is an analog of \eqref{eq:duality_quant}.

A real finite-dimensional matrix $\QSMatr=(\QSMatr_{mn})$ is called \emph{quasistochastic} if all its columns sum to 1: $\sum_m\QSMatr_{mn}=1$ for every $n$.
If moreover all entries are non-negative ($\QSMatr_{mn}\geq0$), then $\QSMatr$ is called \emph{stochastic}.
The norm of $\QSMatr$ is defined as
\begin{equation}
    \Norm{\QSMatr} \Def \max_n \| S_{\bullet n} \|_1,
\end{equation}
where $\QSMatr_{\bullet n}$ denotes the $n$-th column of $\QSMatr$.

\subsection{Frame-based quasiprobability representations} \label{subsec:frame-based_representations}

Roughly speaking, a quasiprobability representation is a way of thinking about quantum mechanics as a statistical mechanics with possible negative probabilities.
In a general context, quasiprobability representations can be introduced using the language of \emph{frames}~\cite{Ferrie_2008, Ferrie_2009, Ferrie_2011, Pashayan_2015, DAriano_2004}.
Certain examples of such representations are discussed in Appendix~\ref{appendix:frame-based}.

We call a frame a pair of linear maps $(\EMap, \eMap)$, where $\EMap : \THil \rightarrow \lOne$ and $\eMap : \lOne \rightarrow \THil$ satisfying
(i) the consistency condition
\begin{equation}~\label{eq:consistency}
	\eMap \circ \EMap = \Id,
\end{equation}
where $\circ$ stands for composition and $\Id$ is the identity map,
(ii) Hermiticity preserving, and (iii) the trace preserving (TP) condition
\begin{equation} \label{eq:TP}
	\sum_m (\EMap[\State])_m = \Tr(\State), \quad \Tr(\eMap[\ProbVecP]) = \sum_m\ProbVecP_m.
\end{equation}
One can see that $\EMap$ translates a quasistate $\State$ as a quasidistribution $\ProbVecP\in\lOne$, and $\eMap$ translates $\ProbVecP$ back as $\State$.
Note that the consistency condition Eq.~\eqref{eq:consistency} implies $M \geq D = d^2$.

A linear map $\EMap$, which is sometimes called an \emph{analysis map}, is defined by a set of $M$ self-adjoint operators $\{\EOp_m\}_{m=1}^M$ ($\EOp_m \in \BHil, \EOp_m=\EOp_m^\dagger$)
that describe the action of $\EMap$ on an input state as follows:
\begin{equation}
	\EMap[\State]= \begin{pmatrix}
		{\Tr}[\State \EOp_1] \\
		\vdots\\
		 {\Tr}[\State \EOp_M]
	\end{pmatrix}.
\end{equation}
Note that the TP property~\eqref{eq:TP} implies
\begin{equation} \label{eq:EsumintoId}
	\sum_m \EOp_m = \Idop.
\end{equation}
Also note that since we require all $\EOp_m$ to be self-adjoint but not necessarily positive semi-definite, $\EMap[\State]$ for a state $\State$ is always a quasidistribution, but not necessarily a distribution.
In the special case where $\EOp_m\geq 0$ for all $m$, the set $\{\EOp_m\}_{m=1}^M$ forms an informationally complete POVM~\cite{Holevo_2019}.

Likewise, a linear operator $\eMap : \lOne \rightarrow \THil$ (sometimes called a \emph{synthesis map}) can be defined by a set of operators $\{\eOp_m\}_{m=1}^M$, where $\eOp_m\in\THil$ for all $m=1,\dots,M$ so that
\begin{equation}
	\eMap[\ProbVecP] = \sum_{m=1}^M \eOp_m \ProbVecP_m =
	\begin{pmatrix}
		\eOp_1 & \ldots & \eOp_M
	\end{pmatrix}
	\begin{pmatrix}
		\ProbVecP_1 \\
		\vdots\\
		\ProbVecP_M
	\end{pmatrix}.
\end{equation}
The TP property implies
\begin{equation} \label{eq:traceone}
	\Tr(\eOp_m)=1.
\end{equation}
In terms of operators, the frame property means that for every $\State \in \THil$
\begin{equation}
	\sum_{m=1}^M \Tr(\State \EOp_m) \eOp_m = \State.
\end{equation}
The frame consistency condition implies \emph{informational completeness}: the linear span of $\{\EOp_m\}_{m=1}^M$ is $\BHil$ and the linear span of $\{\eOp_m\}_{m=1}^M$ is $\THil$.
We call a frame \emph{minimal informationally complete} (MIC) if $M=D$, and informationally complete (IC) -- otherwise.
In the case of MIC frames, $\EMap$ and $\eMap$ are linear isomorphisms of $\THil$ and $\lOne$ correspondingly, so $\{\EOp_m\}_{m=1}^M$ is a linear basis of $\BHil$ and $\{\eOp_m\}_{m=1}^M$ is a dual basis in $\THil$.
Therefore, $\Tr(\eOp_k \EOp_l) = \delta_{k,l}$, where $\delta_{k,l}$ is a standard Kronecker symbol.
We note that the idea of analysis and synthesis maps is close to quantizer-dequantizer approach (see e.g. Refs.~\cite{andreev2020quantizer, man2021probability})

The set $\{1,\dots,M\}$ can be understood as a \emph{phase space}, so that a state $\State$ is ``distributed'' over the phase space via the quasidistribution $\ProbVecP=\EMap[\State]$.
On the other hand, for any given quasidistribution $p\in\lOne$ we can consider a quasistate $\State = \eMap[\ProbVecP]$.
Note that unless $M=D$ (MIC case) the correspondence is not unique:
for a state $\State\in\THil$ there exists a manifold of $\ProbVecP\in\lOne$ such that $\State = \eMap[\ProbVecP]$.
We exploit this phenomenon in Sec.~\ref{sec:general_representations}.

Next, let us consider a quantum channel (CPTP map) $\Chan : \THil \rightarrow \THil$, which describes an evolution of quantum states.
In terms of a quasiprobability representation $(\EMap,\eMap)$, its action can be described by $M\times M$ quasistochastic matrix $\QSMatr = (S_{kl}): \lOne\rightarrow \lOne$ with  elements
\begin{equation}
  \QSMatr_{kl} = \Tr(\EOp_k \Chan[\eOp_l]).
\end{equation}
The fact that it is a quasistochastic is ensured by Eqs.~\eqref{eq:EsumintoId} and \eqref{eq:traceone}, and TP property of $\Chan$.
An action of $\Chan$ then can be represented as an action of quasistochastic matrix $\QSMatr$ on a quasidistribution $\ProbVecP=\EMap[\State]$:
\begin{equation}
	\Chan[\State] = \eMap[\QSMatr\ProbVecP].
\end{equation}
Given a quasistochastic matrix $\QSMatr$, the result of applying the channel to a given state $\State$ can be reconstructed as follows:
\begin{equation} \label{eq:ProperS}
	\Chan[\State] = \sum_{k,l=1}^M \eOp_k \QSMatr_{kl} \Tr(\EOp_l \State).
\end{equation}
We again note that in the overcomplete case $M>D$, there is a manifold of quasistochastic matrices $\QSMatr$ satisfying~\eqref{eq:ProperS}.

To complete our description of quasiprobability representations, consider a measurement given by a POVM $\{\POVMEff_k\}_{k=1}^K$.
In the quasiprobability representation it can be described by a quasistochastic $K\times M$ matrix $\QSMatrM: \lOne^M \rightarrow \lOne^K$ with elements
\begin{equation}
      \QSMatrM_{kl} = {\Tr}[\POVMEff_k \eOp_l].
\end{equation}
One can see that the action of $\QSMatrM$ on a quasidistribution $\ProbVecP=\EMap[\State]$ obtained from a state $\State$ provides a fair probability distribution of measurement outcomes:
\begin{equation} \label{eq:MeasProbs}
	\QSMatrM \ProbVecP=
	\begin{pmatrix}
		{\Tr}[\POVMEff_1\State]\\
		\vdots\\
		{\Tr}[\POVMEff_K\State]
	\end{pmatrix}.
\end{equation}
Note that the $k$-th row of $\QSMatrM$ corresponds to a particular effect $\POVMEff_k$ and can be considered as an element of $\lInf$.

The considered frame-based description of quasiprobability representations can be easily generalized to a multiparticle case where the state space is given by a tensor product $\Hil=\bigotimes_k \Hil^{(k)}$ with $\Hil^{(k)}$ corresponding to a distinct particle (degree of freedom).
In this case, one can consider frame operators of the form
\begin{equation} \label{eq:multiparticle_case}
	\EOp_{\boldsymbol{k}} = \bigotimes_l \EOp_{k_l}^{(l)}, \quad \eOp_{\boldsymbol{k}} = \bigotimes_l \eOp_{k_l}^{(l)},
\end{equation}
where $\{\EOp^{(l)}_k\}_k$ and $\{\eOp^{(l)}_k\}_k$ provide frame operators for each space $\Hil^{(l)}$, and $\boldsymbol{k}=(k_l)$ stands for a multi-index.
Note that Eq.~\eqref{eq:multiparticle_case} describes a \emph{factorizable representation} that is a particular case of possible multiparticle representations.
A further simplification can be obtained by considering the same frame operators ($\EOp_k^{(l)}=\EOp_k$, $\eOp_k^{(l)}=\eOp_k$) for all particles, given that their state spaces are of the same dimension.

\subsection{The role of the negativity in simulation of quantum dynamics}

From the viewpoint of quasiprobability representation, the only difference between classical stochastic dynamics of random variable and quantum dynamics is the presence of negative elements.
Indeed, suppose we fix a frame-based quasiprobability representation $(\EMap,\eMap)$ of a quantum system $\THil$.
If the representation of an element is positive, it is possible to interpret this element using classical statistical mechanics: 
we can understand quantum states as probability distributions over the phase space $\{1,\dots,M\}$ and the action of a channel as a stochastic update of distributions. 
However, it turns out that quasiprobability representations of states, channels and measurements usually have negative entries~\cite{Ferrie_2008}. 
Meanwhile, the more representation is negative (an explicit quantitative measure of negativity is provided further), the more difficult is a simulation of a given circuit using the Monte Carlo algorithm presented in Refs.~\cite{Pashayan_2015, Koukoulekidis_2022}.

For the sake of self-containing, we provide the following algorithm.
Let us consider a state $\State$ passing through a sequence of $L$ quantum channels $\Chan^{(1)}, \ldots, \Chan^{(L)}$ and finally measured with a POVM consisting of a particular effect $\POVMEff$.
The resulting output probability is given by
\begin{equation}
	\OutProb = \Tr(\POVMEff\,\Chan^{(L)}\circ\ldots\circ\Chan^{(1)}[\State]).
\end{equation}
Provided that in some frame $(\EMap,\eMap)$ the state $\State$ is described by a quasidistribution $\ProbVecP$, each channel $\Chan^{(l)}$ by quasistochastic matrix $\QSMatr^{(l)}$, and $\POVMEff$ by row vector $\ProbVecV$,
the output probability $\OutProb$ can be also calculated as follows:
\begin{multline}
	\OutProb = \ProbVecV\,\QSMatr^{(L)} \ldots \QSMatr^{(1)}\ProbVecP=
	\sum_{\Traj}\ProbVecV_{x_{L}} \QSMatr^{(L)}_{x_{L}x_{L-1}} \ldots \QSMatr^{(1)}_{x_1x_0} p_{x_0},
\end{multline}
where $\boldsymbol x=(x_{0},\ldots,x_{L})$ with $x_l\in\{1,\ldots,M\}$ denote possible ``trajectories'' of the state in the phase space during its evolution.
Noticing that elements of employed quasidistributions can be rewritten as
\begin{equation}
	\begin{aligned}
        \ProbVecP_x &=  \mathrm{Sign}(\ProbVecP_x) \|\ProbVecP\|_1 \,\ProbVecP_x^\Stoch, \quad \ProbVecP_x^\Stoch\Def\frac{|\ProbVecP_x|}{\|\ProbVecP\|_1},\\
		\QSMatr^{(l)}_{xy} &= \mathrm{Sign}(\QSMatr^{(l)}_{xy}) \| \QSMatr^{(l)}_{\bullet y}\|_1\,\QSMatr^{(l)\Stoch}_{xy}, \quad \QSMatr^{(l)\Stoch}_{xy} \Def \frac{|\QSMatr^{(l)}_{xy}|}{\| \QSMatr^{(l)}_{\bullet y}\|_1}, \\
	\end{aligned}
\end{equation}
where $\ProbVecP^\Stoch\Def(p_x^\Stoch)_x$ appears to be a fair distribution and $\QSMatr^{(l){\Stoch}}\Def(\QSMatr^{(l)}_{xy})_{xy}$ appear to be stochastic matrices,
one can see that $\OutProb$ can be expressed as an expectation value
\begin{equation}
	\OutProb = \mathbb{E}_{\Traj} \chi(\Traj)
\end{equation}
taken over possible trajectories $\Traj$ of the quantity
\begin{multline}
	\chi(\Traj) =
    \mathrm{Sign}\left(\ProbVecV_{x_L} \prod_{l=1}^L \QSMatr^{(l)}_{l_i,x_{l-1}} \ProbVecP_{x_0}\right) \\
	\times |v_{x_L}| \prod_{l=1}^L\|\QSMatr^{(l)}_{\bullet x_{l-1}}\|_1 \|\ProbVecP\|_1,
\end{multline}
provided that the probability of realizing a trajectory $\Traj$ is given by the Markov-chain expression
\begin{equation} \label{eq:ProbDistrTraj}
	\Pr(\Traj) = \QSMatr^{(L)\Stoch}_{x_{L}x_{L-1}} \ldots \QSMatr^{(1)\Stoch}_{x_1x_0} \ProbVecP^{\Stoch}_{x_0}.
\end{equation}
The value of $\OutProb$ then can be estimated as an average over trajectories sampled according to a random walk process, which is governed by $\ProbVecP^\Stoch$ and $\QSMatr^{(l){\Stoch}}$:
\begin{equation}
	\OutProb \approx \OutProbEst= \frac{1}{N_{\rm samp}} \sum_{\Traj \in \TrajSet } \chi(\Traj),
\end{equation}
where $\TrajSet=\{ \Traj^{(j)}\}_{j=1}^{N_{\rm samp}}$ is a set of $N_{\rm samp}$ trajectories sampled according to~\eqref{eq:ProbDistrTraj}.
An accuracy of the estimation is determined by the absolute value of $\chi(\boldsymbol x)$, which can be uniformly upper bounded as
\begin{equation}
	|\chi(\Traj)| = |\ProbVecV_{x_{L}}|\prod_{l=1}^L\|\QSMatr^{(l)}_{\bullet x_{l-1}}\|_1 \|\ProbVecP\|_1 \leq e^{\NegTot},
\end{equation}
where  $\NegTot$ is the \emph{total logarithmic negativity}
\begin{equation}
	\NegTot = \Neg(\ProbVecP)+\sum_{l=1}^L \Neg(\QSMatr^{(l)}) + \Neg(\ProbVecV),
\end{equation}
that is the sum of \emph{logarithmic negativities} defined as
\begin{equation}
  \begin{aligned}
    \Neg(q) &= \log_d\Norm{q}_1 = \log_d \sum_{m} \Abs{q_m}, \\
	\Neg(\QSMatr) &= \log_d {\|} \QSMatr {\|} = \max_{m} \Neg(\QSMatr_{\bullet m}), \\
    \Neg(w) &= \log_d \Norm{w}_\infty = \log_d \max_m |w_m|,
  \end{aligned}
\end{equation}
for arbitrary quasidistribution $q\in\lOne$, quasistochastic matrix $\QSMatr:\lOne\rightarrow\lOne$, and vector $w\in\lInf$.
While considering a system consisting of $d$-dimensional particles, it is natural to take $\log$ base $d$, e.g. $\log_2$ for multiqubit systems.
We note that in the context of Wigner functions, the logarithmic negativity is usually called \emph{mana} \cite{Veitch_2014}.
We also note that this quantity has a number of nice properties (see Ref.~\cite{Veitch_2014}).
Mainly, it respects the composition of arbitrary quasistochastic matrices $\widetilde{\QSMatr}^{(1)}$ and $\widetilde{\QSMatr}^{(2)}$:
\begin{equation} \label{eq:neg_of_join}
  	\Neg(\QSMatr^{(a)}\circ \QSMatr^{(b)}) \leq \Neg(\QSMatr^{(a)}) + \Neg(\QSMatr^{(b)})
\end{equation}
and is additive with respect to tensor product:
\begin{equation}
  \Neg(\QSMatr^{(a)} \otimes \QSMatr^{(b)}) = \Neg(\QSMatr^{(a)}) + \Neg(\QSMatr^{(b)}).
\end{equation}
One could also inspect the \emph{sum negativity} \cite{Veitch_2014} of quasidistributions, defined as an absolute value of the sum of all negative entries, but it is not very convenient for our needs.

Using the Hoeffding inequality, we get
\begin{equation}
	\Pr[ |\OutProbEst -\OutProb  |\geq \epsilon] \leq 2 \exp\left(- \frac{\epsilon^2N_{\rm samp}}{2\exp({2\NegTot})} \right),
\end{equation}
where $\epsilon$ is a precision parameter.
Thus, to achieve a precision $\epsilon$ with probability at least $1-\delta$, one requires at least
\begin{equation}
	N_{\rm samp} \geq e^{2\NegTot} \frac{2}{\epsilon^2}\ln\frac{2}{\delta}
\end{equation}
samples.
In this way, the total negativity drastically affects the performance of Monte Carlo simulation.
Below, we consider the problem of minimizing $\NegTot$ by exploiting non-uniqueness of quasi-stochastic matrices and quasidistributions for overcomplete frames.

\section{Constructing generalized quasiprobability representations} \label{sec:general_representations}

\subsection{Generalized quasiprobability representations}

In the previous section, we have discussed frame-based quasiprobability representations of quantum mechanics based on pairs of analysis and synthesis maps $(\EMap,\eMap)$.
As we will see below, to construct a quasiprobability representation, it is enough to consider only a synthesis map $\eMap$.
This leads to a notion of \emph{generalized quasiprobability representation}, which is somewhat less restrictive
and allows one to implement an optimization of the negativity in the representation of states and channels.
Some examples (known, yet not considered from our perspective before) of such representations are discussed in Appendix~\ref{appendix:generalized}.

The basic idea behind discarding $\EMap$ from a consideration is as follows.
In the Sec.~\ref{subsec:frame-based_representations} we have noticed that in the overcomplete case $M>D$, distinct quasidistributions $p^{(1)}\neq p^{(2)}$ can correspond to the same quasistate $\State=\eMap[p^{(1)}]=\eMap[p^{(2)}]$.
In fact, there is an infinite number of distinct quasidistributions $p$ satisfying $\eMap[p]=\State$, which form an affine manifold in $\lOne$.
We note that $\EMap[\State]$ is a particular instance of this manifold.
However, these quasidistributions can have different negativity, and we may want to choose the one with the minimal negativity.
The same consideration can be applied to a set quasistochastic matrices that realize the same CPTP map.

In this way, we define a generalized quasiprobability representation as a representation specified only by a map $\eMap : \lOne^M \rightarrow \THil$.
As before, we require $\eMap$, given by a set of operators $\{\eOp_m\}_{m=1}^M$, to satisfy TP ($\Tr\eOp_m=1$) and IC (${\rm Span}\{\eOp_1,\ldots,\eOp_M\}=\THil$) conditions.
One can view $\eMap$ as a \emph{cover} of the space of quantum states $\THil$ by a space of classical quasistates $\lOne$.
In the Heisenberg picture, this corresponds to an \emph{inclusion} of the space of quantum observables $\BHil$ into the space of classical observables $\lInf^M$.
Surely, in the MIC case $M=D$, there exists a unique inverse map $\EMap=\eMap^{-1}$~\cite{Yashin_2020}.
In what follows, we say that we deal with a representation $\eMap$, meaning that we consider a quasiprobability representation defined by $\eMap$.

\begin{figure*}
  \includegraphics[width=0.9\linewidth]{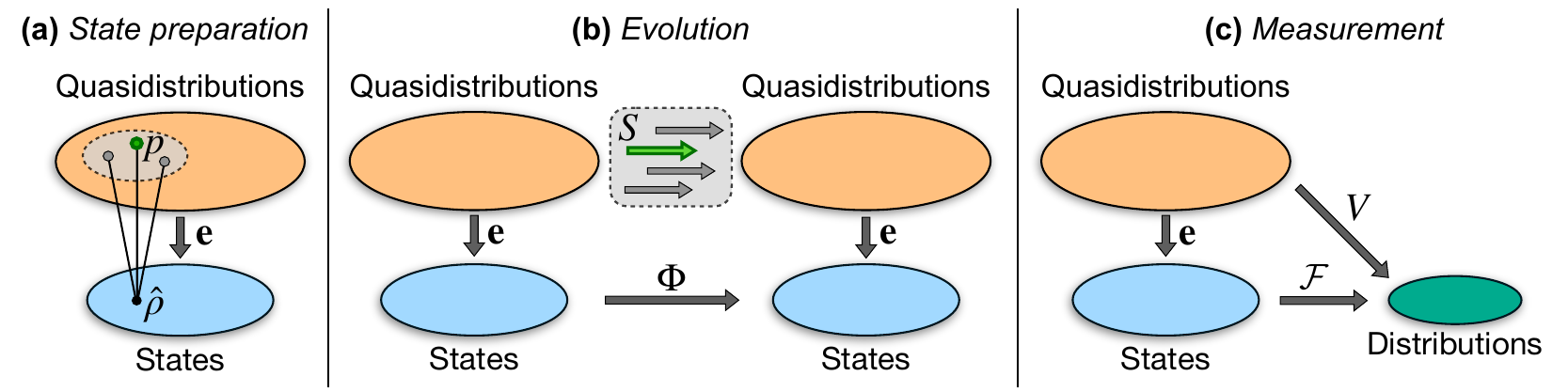}
  \caption{
  The idea behind obtaining states (a), channels (b), and measurements (c) in an overcomplete generalized quasistochastic representation defined by a synthesis map $\eMap$.
  (a) A quantum state $\State$ is represented by a quasidistribution that has the minimal negativity among all quasidistributions $\ProbVecP$ satisfying $\eMap[\ProbVecP]=\State$.
  (b) A quantum channel $\Chan$ is represented by a quasistochastic matrix that has the minimal negativity among all quasistochastic matrices $S$ satisfying $\eMap \circ \QSMatr = \Chan \circ \eMap$.
  (c) A POVM $\POVM$ is represented by a quasistochastic matrix $V$ realizing the map $\POVM \circ \eMap$, which transforms quasidistributions of quantum states into fair distribution of measurement outcomes.
  }
  \label{fig:rep_elements}
\end{figure*}

\subsection{Representation of basic elements}

Here we consider how the basic elements of quantum circuits, namely, states, channels, and measurements, are obtained within the generalized quasiprobability representation to have the minimal possible negativity. 
The synopsis of this analysis is presented in Fig.~\ref{fig:rep_elements}.

\subsubsection{States} \label{subsec:states}

We say that a quasidistribution $p\in\lOne$ represents a state $\State\in\THil$ under $\eMap$ if $\State = \eMap[p]$ [see also Fig.~\ref{fig:rep_elements}(a)].
Let us define a negativity $\Neg(\eMap,\State)$ of a state $\State$ in representation $\eMap$ as the minimal obtainable negativity of all quasidistributions representing $\State$ under $\eMap$:
\begin{equation}
    \Neg(\eMap,\State) \Def \min_{\ProbVecP : \eMap[\ProbVecP] = \State} \Neg(\ProbVecP) = \log \min_{\ProbVecP : \eMap[\ProbVecP] = \State} \Norm{\ProbVecP}_1.
\end{equation}

Let us write the problem of finding $\ProbVecP$ that minimizes negativity for $\State$ as
\begin{equation}
\begin{array}{l l}
  \underset{p}{\text{minimize}}       & \Norm{p}_1 \\
  \text{subject to} & \eMap[p] = \State. \\
\end{array}
\end{equation}
This problem has $M$ real parameters and $\leq D$ constraints.
Note that the constraint $\eMap[p] = \State$ automatically implies $\sum_{m=1}^M p_m = 1$.
Let us use some basic tricks of linear programming (see Ref.~\cite{Boyd_2004}), to rewrite this convex problem as a linear program.
This can be done by introducing additional parameters.
We write a quasidistribution $\ProbVecP$ as a subtraction of positive and negative parts $\ProbVecP = \ProbVecP^+ - \ProbVecP^-$, where $\ProbVecP^\pm \geq 0$ 
(here and hereafter by $v\geq 0$ for an arbitrary $v\in\lOne$ we mean $v_i\geq 0$ for each element of $v$).
Then, one can see that
\begin{equation}
  \Norm{\ProbVecP}_1 = \sum_{m=1}^M (\ProbVecP^+_m + \ProbVecP^-_m)
\end{equation}
In terms of parameters $(\ProbVecP^+,\ProbVecP^-)$, the program can be rewritten as
\begin{equation}
\begin{array}{l l}
  \underset{\ProbVecP^+, \ProbVecP^-}{\text{minimize}}       & \sum_m (p^+_m + p^-_m) \\
  \text{subject to}     & \eMap[\ProbVecP^+] - \eMap[\ProbVecP^-] = \State, \\
                        & \ProbVecP^+ \geq 0, \quad \ProbVecP^- \geq 0.
\end{array}
\end{equation}
This is a linear program with $2M$ parameters, $\leq D$ equality constraints, and $2M$ inequality constraints.
To solve this problem, one can use a highly efficient interior point method realized in the SciPy Python library~\cite{Virtanen_2020}.
We note that similar problems were earlier considered in the context of finding the \emph{robustness of magic}~\cite{Howard_2017, Heinrich_2019}, which is a negativity in the frame of all stabilizer states.

\subsubsection{Channels} \label{subsec:channels}

We say that the quasistochastic matrix $\QSMatr: \lOne \rightarrow \lOne$ represents a channel $\Chan : \THil \rightarrow \THil$ under $\eMap$ if $\eMap\circ \QSMatr = \Chan\circ \eMap$.
This condition on $\QSMatr$ is sometimes called a \emph{co-extension} of $\Chan$ \cite{Yashin_2022}.
That precisely means that if $\ProbVecP\in\lOne$ represents $\State\in\THil$, then the quasidistribution $\QSMatr \ProbVecP \in\lOne$ represents $\Chan[\State]$.

Let us define a negativity $\Neg(\eMap,\QSMatr)$ of a channel $\Chan$ in a representation $\eMap$ as the minimal possible negativity of $\QSMatr$ that represents $\Chan$:
\begin{equation}
	\Neg(\eMap,\Chan) \Def \min_{S: \eMap\circ \QSMatr = \Chan\circ \eMap} \Neg(\QSMatr) = \log \min_{\QSMatr: \eMap\circ \QSMatr = \Chan\circ \eMap} \Norm{\QSMatr}
\end{equation}
[see also Fig.~\ref{fig:rep_elements}(b)].
We note that if one has two channels $\Chan^{(1)}$ and $\Chan^{(2)}$, with corresponding representations $\QSMatr^{(1)}$ and $\QSMatr^{(2)}$, then the quasistochastic matrix $\QSMatr^{(2)}\QSMatr^{(1)}$ represents (usually not optimally) the channel $\Chan^{(2)}\circ \Chan^{(1)}$.
Therefore, it holds that
\begin{equation}
	\Neg(\eMap,\Chan^{(2)}\circ\Chan^{(1)}) \leq \Neg(\eMap,\Chan^{(1)}) + \Neg(\eMap,\Chan^{(2)}).
\end{equation}

The problem of finding the representation $\QSMatr$ of a channel $\Chan$ with the minimal negativity can be stated as
\begin{equation}
\begin{array}{l l}
	\underset{\QSMatr}{\text{minimize}}       & \Norm{\QSMatr} \\
	\text{subject to}     & \eMap \circ \QSMatr = \Chan \circ \eMap. \\
\end{array}
\end{equation}
Since $\Norm{S} = \max_l \Norm{S_{\bullet l}}_1$ is a maximum over all columns, in order to find the representation of $\Chan$ with the minimal negativity it is sufficient to optimize the representation on each column:
\begin{equation}
\begin{aligned}
	\Neg(\eMap,\Chan)
	&= \log \min_{S: \eMap\circ S = \Chan\circ \eMap} \Norm{S} \\
	&= \max_{m=1\dots M} \Neg(\eMap,\Chan[\eOp_m]).
\end{aligned}
\end{equation}
Thus, we reduce the problem to finding the optimal representations for $M$ quasistates $\Chan[\eOp_m]$ for $m=1,\dots,M$.
We note that solving this problem can be parallelized, and in the worst case, it is $M$ times slower than solving the corresponding problem for a single state.

\subsubsection{Measurements} \label{subsec:measurements}

Suppose we are given a POVM $\POVM : \THil \rightarrow \lOne^K$ represented by operators $\{\POVMEff_k\}_{k=1}^K$; the action of $\POVM$ on a state $\State$ is defined as $\POVM[\State]=\begin{pmatrix} {\Tr}[\POVMEff_1\State]& \ldots& {\Tr}[\POVMEff_K\State] \end{pmatrix}^{\sf T}$.
From the viewpoint of a generalized quasistochastic representation, there is only one representation of this measurement with a quasistochastic matrix $\QSMatrM$ representing the map $\POVM\circ \eMap$, i.e. $\QSMatrM_{kl} = {\Tr}(\POVMEff_k \eOp_l)$ [see also Fig.~\ref{fig:rep_elements}(c)].
The negativity of a measurement is defined as
\begin{equation}
  \Neg(\eMap,\POVM) = \log\Norm{\POVM\circ \eMap} = \log\max_{l=1\dots M} \norm{V_{\bullet l}}_1.
\end{equation}
In contrast to states and channels, no optimization problems are required to be solved.
Note that for a trivial POVM $\POVM=\{\hat{\mathbb{1}}\}$, one has
$V=\begin{pmatrix}
    1 & \ldots & 1
\end{pmatrix}$
with zero negativity.

\subsection{Extending the generalized representation to the multiparticle case} \label{subsec:tensor_product}

For a multiparticle case, i.e., when $\Hil=\bigotimes_k\Hil^{(k)}$ (here $\Hil^{(k)}$ is the state space of each particular particle), we can consider three approaches of how the generalized representation can be constructed.
The first one is to consider some general map $\eMap: \bigotimes_k\Hil^{(k)} \rightarrow \lOne^M$, that transforms a joint state of all particles to some quasiprobability distribution.
One can see that the complexity of defining $\eMap$ in this general case is exponential in the number of particles.

The second approach is to split the joint space into subspaces corresponding to distinct particles or disjoint groups of particles (e.g., pairs or triples), and consider a synthesis map in the form  $\eMap = \bigotimes_k \eMap^{(k)}$.
By this we mean that
\begin{equation} \label{eq:fatorized-repr}
    \eOp_{\boldsymbol{j}} = \bigotimes_k \eOp^{(k)}_{j_k},
\end{equation}
where $\boldsymbol{j}=(j_k)_k$ is a multiindex.
The complexity of specifying $\eMap$ in this case is linear in the number of groups.
However, the introduced restriction on the form of frame operators in general increases the negativity.
Note that the negativity of tensor products turns into sums of negativity of each factor:
\begin{equation}
    \begin{aligned}
      &\Neg(\eMap^{(a)}\otimes \eMap^{(b)},\State^{(a)}\otimes \State^{(b)}) = \sum_{l=a,b}\Neg(\eMap^{(l)},\State^{(l)}),\\
      &\Neg(\eMap^{(a)}\otimes \eMap^{(b)},\Chan^{(a)}\otimes \Chan^{(b)}) =\sum_{l=a,b} \Neg(\eMap^{(l)},\Chan^{(l)}),
    \end{aligned}
\end{equation}
where $\eMap^{(l)}$, $\State^{(l)}$, and $\Chan^{(l)}$ ($l\in\{a,b\}$) are some synthesis maps, states, and CPTP maps correspondingly.

One of the simplest approaches for representation construction is considering the same map $\eMap^{(k)}=\eMap^{(1)}$ for each single particle.
In this case,
\begin{equation} \label{eq:e1q}
    \eOp_{\boldsymbol{j}} = \bigotimes_k \eOp^{(1)}_{j_k}.
\end{equation}
Of course, this approach works for particles of the same dimensionality (e.g., a multiqubit system).
Although it may provide non-optimal negativity for considered operations, it is very suitable for numerical optimization, considered in the next section.

\section{Searching for the optimal representation for a given quantum circuit}\label{sec:optimization}

In the previous section we have considered the problem of negativity minimization with respect to freedom appearing due to an overcompleteness of a given generalized frame-based representation.
Here, we consider the next problem of minimizing negativity for a given circuit by considering a joint optimization with respect to possible maps $\eMap$ and with respect to particular representations of the circuit's basic elements, specified by $\eMap$.

\begin{figure}
    \centering
    \includegraphics[width=\linewidth]{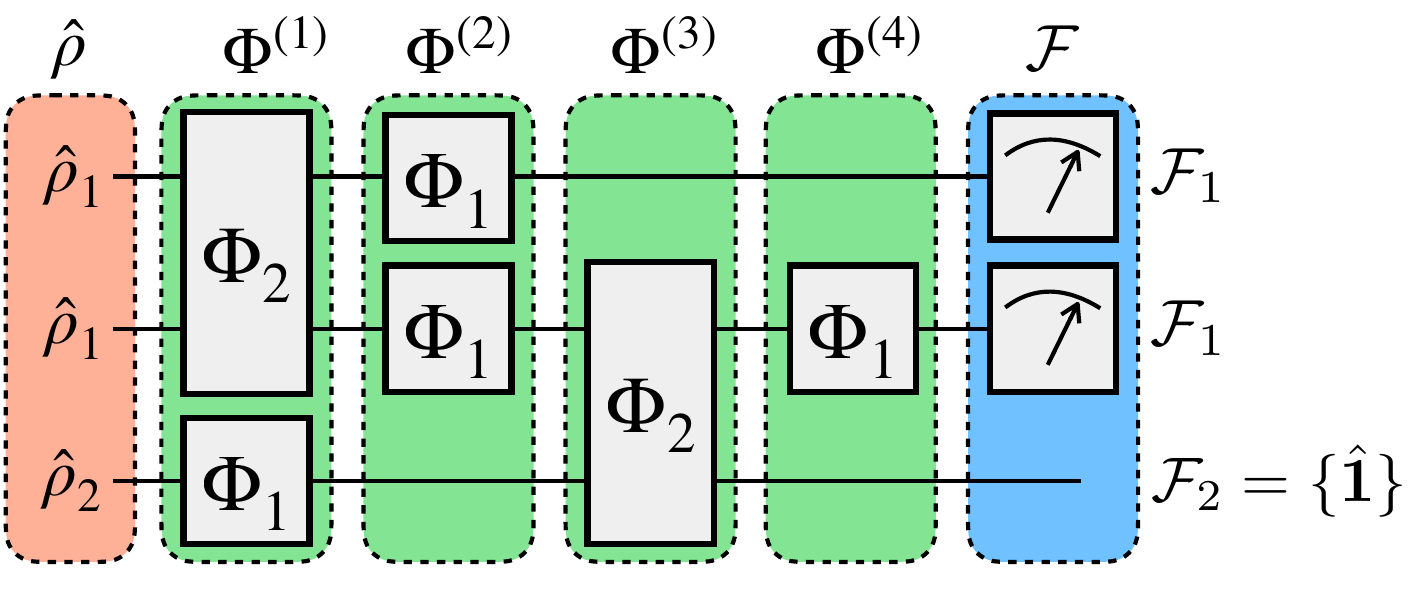}
    \caption{
    An example of a three-qudit circuit $\Circ$, where $\State_1$ and $\State_2$ are some single-qudit states, $\Phi_1$ and $\Phi_2$ are respectively some single- and two-qudit channels, $\POVM_1$ is some single-qudit measurement, and $\POVM_2=\{\hat{\mathbb{1}}\}$ stands for an absence of measurement.
    }
    \label{fig:rep_circuit}
\end{figure}

Consider a circuit $\Circ$ on a set of $n$ qudits ($d$-level quantum particles).
For simplicity, we assume that the circuit does not contain measurement-based feedback controls.
Formally, $\Circ$ can be defined as a triple $(\State, (\Chan^{(k)})_k, \POVM)$, where $\State$ is the initial state of qudits' register, $(\Chan^{(l)})_l$ is a sequence of quantum channels, acting on the initial state, and $\POVM$ is a POVM of read-out measurement (see Fig.~\ref{fig:rep_circuit}).
Given a synthesis map $\eMap: \THil \rightarrow \lOne^M$ with $M\geq d^{2n}$, it is natural to define the negativity of the circuit in the quasiprobability representation as
\begin{equation}
  \Neg(\eMap,\Circ) = \Neg(\eMap,\State) + \sum_{l=1}^K \Neg(\eMap,\Chan^{(l)}) + \Neg(\eMap,\POVM).
\end{equation}
Note that defining negativity as a logarithm of norms is convenient, because it is additive with respect to compositions.
The product of norms of all elements rises exponentially with the increase of number of elements $K$, while the logarithmic negativity rises linearly.

The problem of minimizing the negativity for $\Circ$ then takes the form
\begin{equation} \label{eq:fulloptprob}
\begin{array}{l l}
  \underset{\eMap\in {\cal E}}{\text{minimize}}       & \Neg(\eMap,\Circ), \\
\end{array}
\end{equation}
where ${\cal E}$ is a set of considered synthesis operators discussed next.

Since the only constraint on the operators $\{\eOp_k\}$, which realize $\eMap$, is the TP property, in the most general case the optimization problem has
\begin{equation}
    N_{\rm par} = (d^{2n}-1)M\geq (d^{2n}-1)d^{2n}
\end{equation}
independent unconstrained parameters.
To decrease the number of parameters, one can consider a set ${\cal E}$ consisting of factorized representations of the form~\eqref{eq:fatorized-repr}.
It results in
\begin{equation}
    N_{\rm par} = n(d^{2}-1)\prod_kM_k\geq n(d^{2}-1)d^2
\end{equation}
independent parameters (here $M_k$ is an output dimension for each $\eMap^{(k)}$).

To further reduce the number of parameters, one can also consider the same single-qudit synthesis map $\eMap_{\rm 1q}$ for all qudits [see Eq.~\eqref{eq:e1q}].
It results in
\begin{equation}
    N_{\rm par} = (d^{2}-1)M\geq (d^{2}-1)d^2
\end{equation}
parameters for the optimization problem.
Taking into account that in real experimental setups $\State$ is usually given by a tensor product of states from some set of lower dimensional states $\{\State_\alpha\}$ (e.g., single-qudit states), 
each $\Chan^{(k)}$ realizes a single-qudit or two-qudit operation, i.e., is given by a single-qudit or two-qudit channel taken from some set $\{\Phi_\beta\}$ tensor multiplied by an identity operator of an appropriate dimension, 
and each effect of $\POVM$ is given as a tensor product of single-qudit effects of POVMs from a set $\{\POVM_\gamma\}$ (see an example in Fig.~\ref{fig:rep_circuit}), and we obtain
\begin{equation}
    \begin{aligned}
        \Neg((\eMap^{(1)})^{\otimes n}, \Circ)
        =&\sum_\alpha s_\alpha\Neg((\eMap^{(1)})^{\otimes {\cal D}_\alpha},\State_\alpha) \\
        &+ \sum_\beta s_\beta\Neg((\eMap^{(1)})^{\otimes {\cal D}_\beta},\Chan_\beta) \\
        &+ \sum_\gamma s_\gamma\Neg((\eMap^{(1)})^{\otimes {\cal D}_\gamma},\POVM_\gamma)\\
       \equiv &\Neg(\eMap^{(1)},\Circ),
    \end{aligned}
\end{equation}
where $s_\alpha$, $s_\beta$, and $s_\gamma$ are numbers of times $\State_\alpha$, $\Chan_\beta$, and $\POVM_\gamma$ appear in the circuit $\Circ$ respectively and ${\cal D}_\alpha$, ${\cal D}_\beta$, and ${\cal D}_\gamma$ are numbers of qudits used to describe the corresponding element.

E.g., for the circuit shown in Fig.~\ref{fig:rep_circuit}, one has
\begin{multline}
\Neg(\eMap^{(1)},\Circ)=
2\Neg(\eMap^{(1)},\rho_1)+\Neg(\eMap^{(1)},\rho_2)\\  +4\Neg(\eMap^{(1)},\Phi_1)+2\Neg((\eMap^{(1)})^{\otimes 2},\Phi_2)\\
  +2\Neg(\eMap^{(1)},\POVM_1).
\end{multline}

Unfortunately, the optimization problem~\eqref{eq:fulloptprob} appears to be hard to solve.
The cost function $\Neg(\eMap,\Circ)$ is continuous and locally Lipschitz, but is nonconvex and nonsmooth, that makes it difficult to employ gradient descent-based methods.
In what follows, we consider the simplest case where optimization is performed over a single one-qudit map $\eMap^{(1)}$.
We note that even in this case, the number of degrees of freedom $N_{\rm par}$ varies from tens to hundreds depending on the value of $M$.

As the basic solver for the optimization problem with respect to $\eMap^{(1)}$ we use the SLSQP algorithm.
We apply the optimization algorithm several times, varying initial points and picking the best solution.
This looks like a coarse version of a global optimization.
Nevertheless, it usually gives the same result as global optimizers, such as ``basin hopping'', but is many times faster since it does not require a grid of reference points (about hundreds of points versus tens in our approach).
We note that the global non-convex optimization problem is NP-hard~\cite{jain2017non}, and our approach, as well as other global optimization algorithms, does not guarantee the convergence to the actual global minima.

\section{Performance analysis} \label{sec:performance}

Here, we apply the developed algorithm to some cases of qubit quantum circuits.
We note that we use log base 2 to compute values of negativity.

\subsection{Clifford+T noisy circuits}

Here we apply the negativity minimization procedure for blocks consisting of qubit standard state preparation and measurement (SPAM) operations, Clifford group generators, and T gates.
Recall that the full Clifford group can be generated by single-qubit Hadamard and phase gates
\begin{equation}
    \hat{{\sf H}} \equiv \frac{1}{\sqrt{2}}\begin{pmatrix}
        1 & 1 \\ 1 & -1
    \end{pmatrix},\quad
    \hat{{\sf S}} \equiv \begin{pmatrix}
        1 & 0 \\ 0 & \imath
    \end{pmatrix},
\end{equation}
applicable to each qubit, and two-qubit controlled-NOT gates
\begin{equation}
    \hat{{\sf CX}} \equiv \begin{pmatrix}
        1 & 0 & 0 & 0\\
        0 & 1 & 0 & 0\\
        0 & 0 & 0 & 1\\
        0 & 0 & 1 & 0\\
    \end{pmatrix},
\end{equation}
applicable within some connected graph of couplings between qubits.
Adding a non-Clifford $\pi/8$ phase gate (T gate)
\begin{equation}
    \hat{{\sf T}} \equiv \begin{pmatrix}
        1 & 0 \\ 0 & e^{\imath\pi/4}
    \end{pmatrix},
\end{equation}
makes the gate set universal~\cite{Boykin_1999,Nielsen_2002}.

Let us denote the density matrix of the standard qubit initial state and the computational basis measurement POVM as
$\State_0 \equiv \ket{0}\bra{0}$ and $\\POVM_z \equiv \{\ket{0}\bra{0},\ket{1}\bra{1}\}$ correspondingly.
We will consider primitive circuits consisting of the following elements (the order of elements is insignificant):
\begin{enumerate}[(i)]
  \item single-qubit Clifford block
  \begin{equation} \label{eq:1qcliff}
      \Circ_{\rm1q} = \{\State_0, \hat{{\sf H}}, \hat{{\sf S}}, \POVM_z\},
  \end{equation}
  \item single-qubit Clifford+T block
  \begin{equation}
      \Circ_{\rm1q+T} = \{\State_0, \hat{{\sf H}}, \hat{{\sf S}}, \hat{{\sf T}}, \POVM_z\}
  \end{equation}
  \item two-qubit Clifford block
  \begin{equation}
      \Circ_{\rm2q} = \{\State_0, \hat{{\sf H}}, \hat{{\sf S}}, \hat{{\sf CX}}, \POVM_z\},
  \end{equation}
  \item two-qubit Clifford+T block
  \begin{equation} \label{eq:2qcliffT}
      \Circ_{\rm2q+T} = \{\State_0, \hat{{\sf H}}, \hat{{\sf S}},\hat{{\sf T}}, \hat{{\sf CX}}, \POVM_z\}.
  \end{equation}
\end{enumerate}

\begin{figure*}
    \centering
    \includegraphics[width=0.9\linewidth]{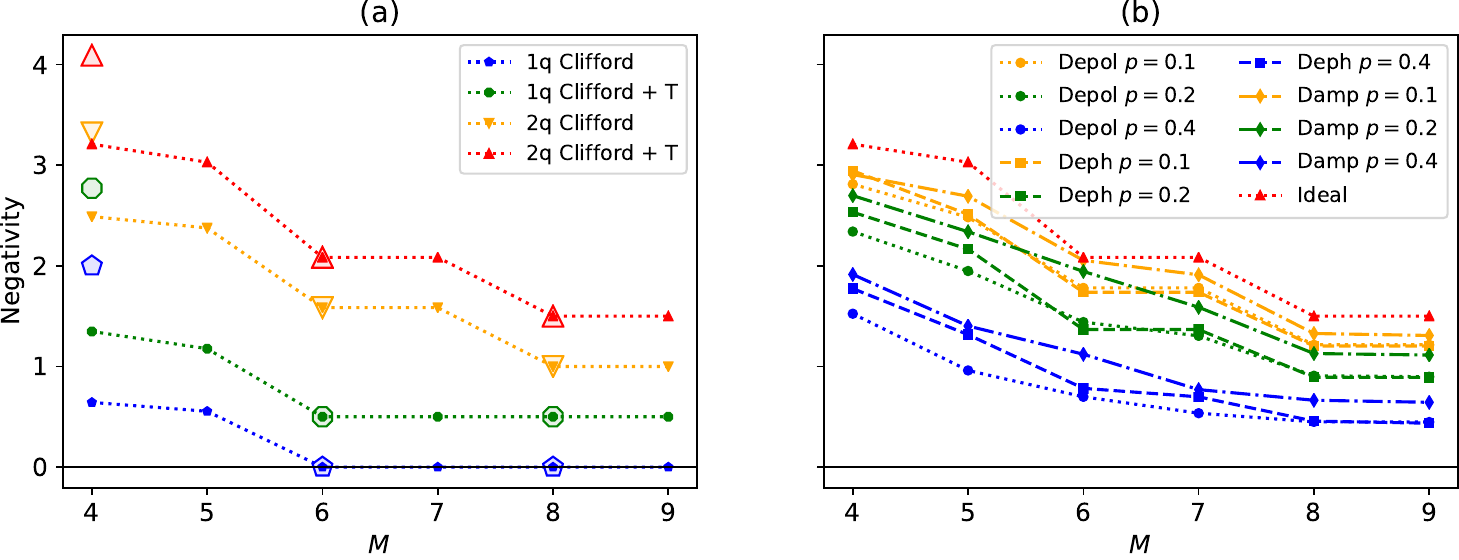}
    \caption{Results of negativity minimization for Clifford+T-based blocks and different frame dimension $M$.
    In (a) the resulting negativity for blocks of the form~\eqref{eq:1qcliff}-\eqref{eq:2qcliffT} is shown.
    Additional empty symbols depict  negativities of corresponding blocks for standard frames (Wootters for $M=4$, stabilizer for $M=6$, and $\Lambda$ polytope for $M=8$).
    In (b) the resulting negativity for the two-qubit Clifford+T block~\eqref{eq:2qcliffT} affected by different types of decoherence channels is presented.
    The obtained values may not coincide with global optima and could probably be improved.
    }
    \label{fig:cliff_circs}
\end{figure*}

The results of negativity minimization with respect to factorized representation consisting of identical synthesis maps on each qubit (see Eq.~\eqref{eq:e1q} for $M=4,\ldots,9$ are shown in Fig.~\ref{fig:cliff_circs}~(a).
We see that the only the block whose negativity drops to zero is $\Circ_{\rm1q}$ for $M\geq 6$, as expected by the Gottesman-Knill theorem~\cite{Nielsen_2002}.
All other blocks, including $\Circ_{\rm2q}$, have nonzero negativities for all values of $M$.
Next, we note that negativity values reach a plateau at $M=6$ for single-qubit blocks $\Circ_{\rm1q}$ and $\Circ_{\rm1q+T}$, and at $M=8$ for two-qubit blocks $\Circ_{\rm2q}$ and $\Circ_{\rm2q+T}$.
The results of numerical optimization [not presented in Fig.~\ref{fig:cliff_circs}(a)] show that negativity remains the same for larger values of $M$.
In Fig.~\ref{fig:cliff_circs}(a), we also depict points corresponding to negativities with respect to some standard representations 
[see Appendix~\ref{appendix:frame-based} and~\ref{appendix:generalized}], 
namely, Wootters ($M=4$), stabilizer states based $(M=6)$, and $\Lambda$ polytope ($M=8$).
In fact, the developed optimization algorithm rediscovers stabilizer frame and $\Lambda$ polytope, which provide the minimal negativity at corresponding values of $M$. 
Yet, for $M=4$, the optimization procedure finds a representation that is better than Wootters and SIC-POVM based (the latter provides larger negativity than the Wootters representation).

We then study an effect of decoherence on negativity of $\Circ_{\rm2q+T}$.
We consider depolarization (depol), dephasing (deph), and amplitude damping (damp) channels given respectively by the following Kraus  operators:
\begin{equation} \label{eq:dec-channels}
    \begin{aligned}
        &\hat A_{0}^{\rm depol}(p)=\sqrt{1-\frac{3p}{4}}\hat{\mathbb{1}},  &&\hat A_{i}^{\rm depol}(p)= \sqrt{\frac{p}{4}}\hat\sigma_i,\\
        &\hat A_{1}^{\rm deph}(p) =\sqrt{1-\frac{p}{2}} \hat{\mathbb{1}},  &&\hat A_{2}^{\rm deph}(p)= \sqrt{\frac{p}{2}}\hat\sigma_3 \\
        &\hat A_{1}^{\rm damp}(p)=\begin{pmatrix}
            1 & 0\\0 & \sqrt{1-p}
        \end{pmatrix},  &&\hat A_{2}^{\rm damp}(p)=\begin{pmatrix}
            0 & \sqrt{p} \\ 0 & 0
        \end{pmatrix},
    \end{aligned}
\end{equation}
where $i=1,2,3$, $\hat\sigma_i$ stand for standard Pauli matrices, and $p\in[0,1]$ specifies decoherence strength.
We add application of one of these decoherence channels after each of the unitary channels, so the resulting channels of single-qubit ${\hat U}$ and two-qubit ${\hat V}$ gates take respectively the form
\begin{eqnarray}
        & \Phi_{\hat U}({\cal A})[\State] = \sum_i \hat A_i \hat U \State \hat U^\dagger \hat A_i^\dagger, \\
        & \Phi_{\hat V}({\cal A})[\State] = \sum_{i,j} (\hat A_i\otimes \hat A_j) \hat V \State \hat V^\dagger (\hat A_i\otimes \hat A_j)^\dagger, \label{eq:two-qubit_dec}
\end{eqnarray}
where ${\cal A}=\{\hat A_i\}_i$ is a set of decoherence channel Kraus operators.
SPAM operators remain ideal.

The resulting values of negativity for $\Circ_{\rm2q+T}$ with presence of decoherence ($p=0.1,0.2,0.4$)  are shown in Fig.~\ref{fig:cliff_circs}~(b).
As one may expect, the negativity drops with increasing decoherence strength $p$.
A plateau still remains for $M\geq 8$, yet results for $M=7$ become different from the ones for $M=6$ (e.g., see damping at $p=0.2$).
In Table~\ref{tab:cliff} we compare obtained values of the negativity with the ones obtained for standard representations.
One can see that the optimization procedure finds better representations for all values of $M$.
Thus, we conclude that the form of the optimal representation changes when we add decoherence effects into consideration.

\begin{table*}[]
\begin{tabular}{c|ccc|cc|cc|}
\cline{2-8}
                                    & \multicolumn{3}{c|}{$M=4$}                                                     & \multicolumn{2}{c|}{$M=6$}                      & \multicolumn{2}{c|}{$M=8$}                              \\ \cline{2-8}
                                    & \multicolumn{1}{c|}{Wootters} & \multicolumn{1}{c|}{SIC-POVM}  & Optimized           & \multicolumn{1}{c|}{Stabilizers} & Optimized     & \multicolumn{1}{c|}{$\Lambda$-polytope} & Optimized     \\ \hline
\multicolumn{1}{|c|}{Ideal}         & \multicolumn{1}{c|}{4.09}   & \multicolumn{1}{c|}{5.49} & 3.21 ($22\%, 42\%$) & \multicolumn{1}{c|}{2.06}      & 2.06 ($0\%$)  & \multicolumn{1}{c|}{1.5}                & 1.5 ($0\%$)   \\ \hline
\multicolumn{1}{|c|}{Depol $p=0.1$} & \multicolumn{1}{c|}{3.65}    & \multicolumn{1}{c|}{5.02} & 2.81 ($23\%, 44\%$) & \multicolumn{1}{c|}{1.78}       & 1.78 ($0\%$)  & \multicolumn{1}{c|}{1.24}               & 1.22 ($2\%$)  \\ \hline
\multicolumn{1}{|c|}{Depol $p=0.2$} & \multicolumn{1}{c|}{3.17}    & \multicolumn{1}{c|}{4.49} & 2.34 ($26\%, 48\%$) & \multicolumn{1}{c|}{1.44}       & 1.44 ($0\%$)  & \multicolumn{1}{c|}{0.94}               & 0.91 ($4\%$)  \\ \hline
\multicolumn{1}{|c|}{Depol $p=0.4$} & \multicolumn{1}{c|}{2.07}    & \multicolumn{1}{c|}{3.28} & 1.52 ($26\%, 54\%$) & \multicolumn{1}{c|}{0.85}       & 0.7 ($18\%$)  & \multicolumn{1}{c|}{0.49}               & 0.45 ($7\%$)  \\ \hline
\multicolumn{1}{|c|}{Deph $p=0.1$}  & \multicolumn{1}{c|}{3.8}     & \multicolumn{1}{c|}{5.15} & 2.95 ($22\%, 43\%$) & \multicolumn{1}{c|}{1.74}       & 1.74 ($0\%$)  & \multicolumn{1}{c|}{1.27}               & 1.2 ($6\%$)   \\ \hline
\multicolumn{1}{|c|}{Deph $p=0.2$}  & \multicolumn{1}{c|}{3.48}    & \multicolumn{1}{c|}{4.79} & 2.53 ($27\%, 47\%$) & \multicolumn{1}{c|}{1.37}       & 1.37 ($0\%$)  & \multicolumn{1}{c|}{1.03}               & 0.89 ($13\%$) \\ \hline
\multicolumn{1}{|c|}{Deph $p=0.4$}  & \multicolumn{1}{c|}{2.79}    & \multicolumn{1}{c|}{4.0}  & 1.77 ($36\%, 56\%$) & \multicolumn{1}{c|}{0.78}       & 0.78 ($0\%$)  & \multicolumn{1}{c|}{0.68}               & 0.46 ($33\%$) \\ \hline
\multicolumn{1}{|c|}{Damp $p=0.1$}  & \multicolumn{1}{c|}{3.92}    & \multicolumn{1}{c|}{5.23} & 2.91 ($26\%, 44\%$) & \multicolumn{1}{c|}{2.05}       & 2.05 ($0\%$)  & \multicolumn{1}{c|}{1.35}               & 1.33 ($2\%$)  \\ \hline
\multicolumn{1}{|c|}{Damp $p=0.2$}  & \multicolumn{1}{c|}{3.73}    & \multicolumn{1}{c|}{4.94} & 2.7 ($28\%, 45\%$)  & \multicolumn{1}{c|}{2.0}        & 1.95 ($3\%$)  & \multicolumn{1}{c|}{1.18}               & 1.13 ($5\%$)  \\ \hline
\multicolumn{1}{|c|}{Damp $p=0.4$}  & \multicolumn{1}{c|}{3.28}    & \multicolumn{1}{c|}{4.27} & 1.91 ($42\%, 55\%$) & \multicolumn{1}{c|}{1.83}       & 1.12 ($39\%$) & \multicolumn{1}{c|}{0.79}               & 0.66 ($16\%$) \\ \hline
\end{tabular}
\caption{Comparison of negativity values for noisy $\Circ_{\rm2q+T}$ block in standard representations and obtained using the developed optimization algorithm.
In brackets, the relative improvement after optimization is shown.}
\label{tab:cliff}
\end{table*}

\subsection{Noisy variational circuit}

\begin{figure}
    \centering
    \includegraphics[width=0.75\linewidth]{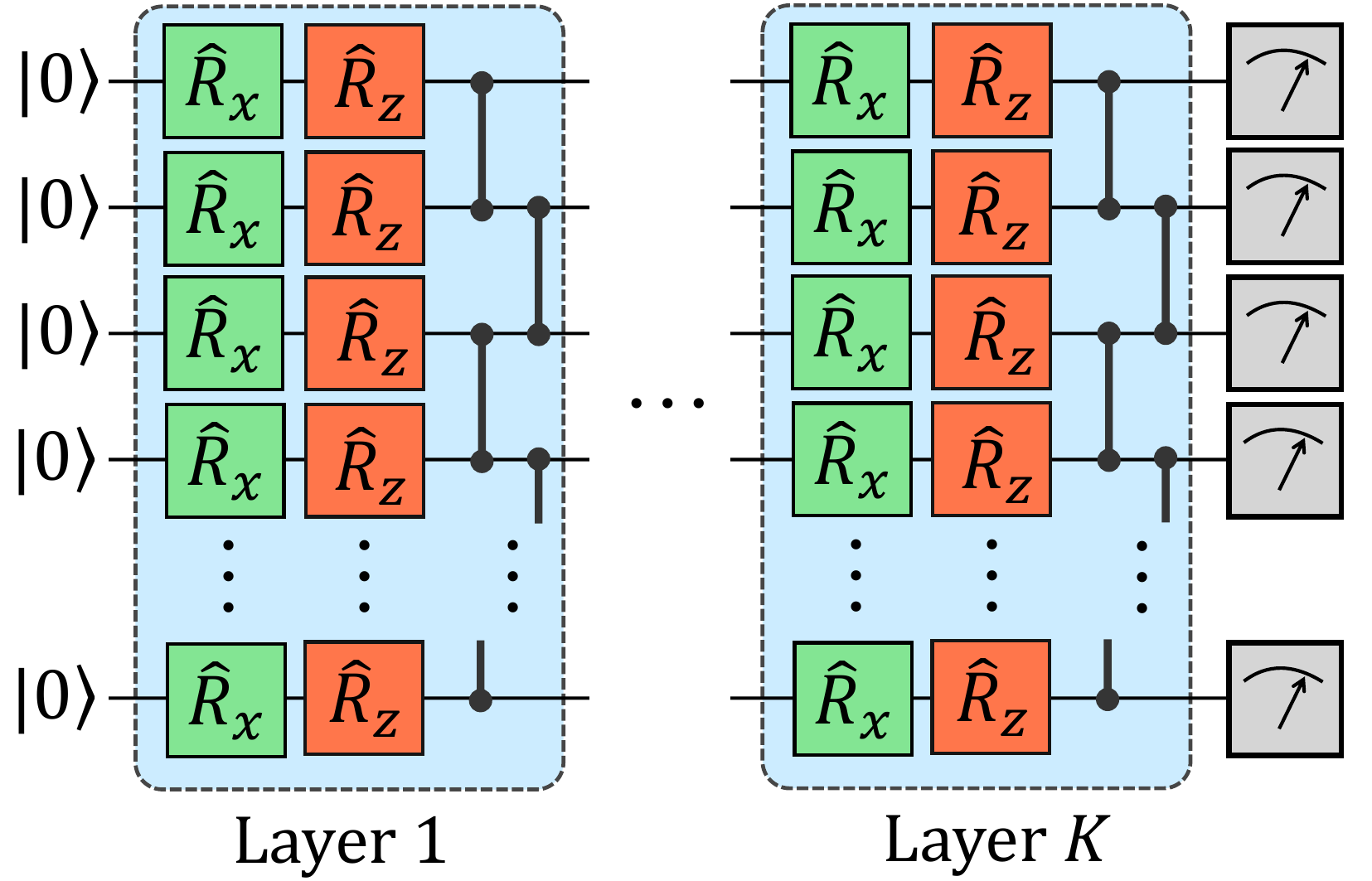}
    \caption{Structure of the considered variational circuit.
    Standard notation for two-qubit controlled phase gates $\hat{\sf CZ}$ is used.
    Angles for single-qubit rotation gates are omitted.
    Read-out measurements are performed in arbitrary single-qubit bases.}
    \label{fig:var_circuit}
\end{figure}

We also apply the negativity minimization algorithm for a variational circuit with layered structure shown in Fig.~\ref{fig:var_circuit}.
Each layer consists of single-qubit rotations
\begin{equation}
    \hat R_x(\theta)=e^{-\imath \hat\sigma_1 \theta/2}, \quad
    \hat R_z(\theta)=e^{-\imath \hat \sigma_3 \theta/2}
\end{equation}
around the $x$ and $z$ axis of the Bloch sphere correspondingly, and two-qubit controlled phase gates
\begin{equation}
    \hat {\sf CZ} = {\rm diag}(1,1,1,-1)
\end{equation}
that couple qubits along one-dimensional chain topology.
In this way, each layer of the $n$-qubit circuit consists of $2n$ single-qubit and $n-1\approx n$ two-qubit gates.

To consider a general case, we take a grid of $B:=10$ points
\begin{equation}
    \Theta = \{-\pi, -\pi+\frac{2\pi}{B}, \ldots, -\pi+\frac{2\pi(B-1)}{B}\}
\end{equation}
on $[-\pi, \pi)$, and consider a set of gates $\Circ_{\rm var}$ consisting of
$\{\hat{R}_x(\theta)\}_{\theta\in\Theta}$, $\{\hat{R}_z(\theta)\}_{\theta\in\Theta}$, and $B$ copies of $\hat{\sf CZ}$ (thus, $\Circ_{\rm var}$ consists of $2B$ single-qubit and $B$ two-qubit gates).
We assume that for a large number of layers the negativity of gates in the layers dominates over negativity of SPAM operations, so SPAM operations are not considered within application of the negativity minimization algorithm.
We additionally consider a noisy version of a gate set by applying the same approach as in the previous subsection.

\begin{figure}
    \centering
    \includegraphics[width=\linewidth]{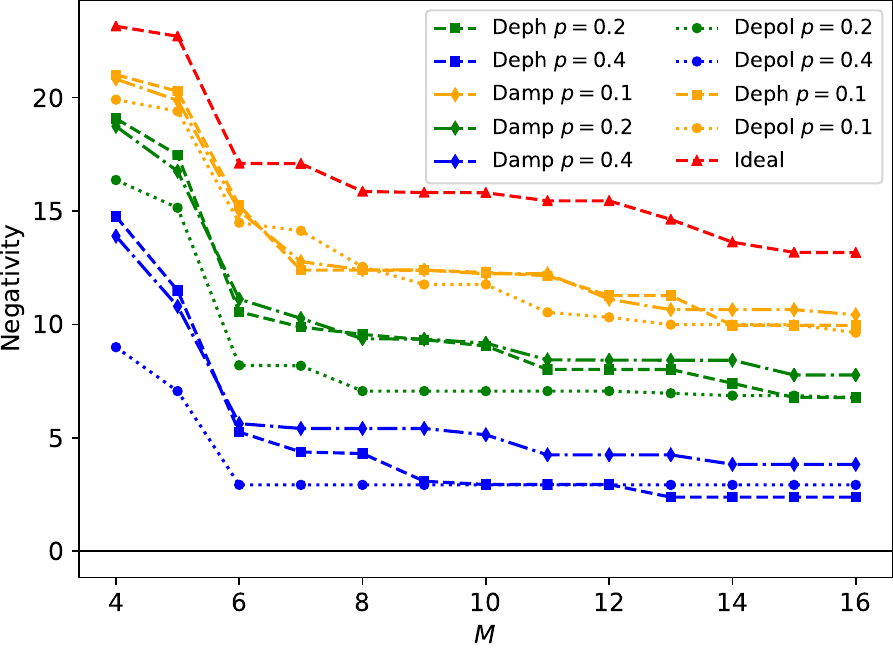}
    \caption{Results of negativity minimization for a variational circuit with respect to different frame dimensionalities $M$ and different types of decoherence.
    The obtained values may not coincide with global optima, and could probably be improved.}
    \label{fig:var_circ_rslts}
\end{figure}

The results of the negativity minimization for $\Circ_{\rm var}$ are shown in Fig.~\ref{fig:var_circ_rslts} (factorized representation consisting of an identical synthesis map is considered).
We can see that, as expected, negativity decreases with growths of representation dimension $M$ and decoherence strength $p$.
In contrast to Clifford+T blocks, a slight possible improvement for $M>8$ is observed, yet the negativity clearly tends to some nonzero limit.

In Table~\ref{tab:variat}, we present the results of comparing the obtained values of the negativity with the ones obtained for standard representations.
One can see the significant improvement after applying the optimization algorithm for all considered values of $M$ and decoherence types.

\begin{table*}[]
\begin{tabular}{c|ccc|cc|cc|}
\cline{2-8}
                                    & \multicolumn{3}{c|}{$M=4$}                                                     & \multicolumn{2}{c|}{$M=6$}                      & \multicolumn{2}{c|}{$M=8$}                              \\ \cline{2-8}
                                    & \multicolumn{1}{c|}{Wootters} & \multicolumn{1}{c|}{SIC-POVM}  & Optimized           & \multicolumn{1}{c|}{Stabilizer} & Optimized     & \multicolumn{1}{c|}{$\Lambda$-polytope} & Optimized     \\ \hline
\multicolumn{1}{|c|}{Ideal}         & \multicolumn{1}{c|}{26.3}    & \multicolumn{1}{c|}{32.3} & 23.1 ($12\%, 28\%$) & \multicolumn{1}{c|}{22.4}       & 17.1 ($24\%$) & \multicolumn{1}{c|}{16.5}               & 15.9 ($4\%$)  \\ \hline
\multicolumn{1}{|c|}{Depol $p=0.1$} & \multicolumn{1}{c|}{23.4}    & \multicolumn{1}{c|}{29.2} & 19.9 ($15\%, 32\%$) & \multicolumn{1}{c|}{18.4}       & 14.5 ($21\%$) & \multicolumn{1}{c|}{15.2}               & 12.4 ($18\%$) \\ \hline
\multicolumn{1}{|c|}{Depol $p=0.2$} & \multicolumn{1}{c|}{20.4}    & \multicolumn{1}{c|}{25.7} & 16.4 ($20\%, 36\%$) & \multicolumn{1}{c|}{14.0}       & 8.2 ($41\%$)  & \multicolumn{1}{c|}{13.8}               & 9.4 ($32\%$)  \\ \hline
\multicolumn{1}{|c|}{Depol $p=0.4$} & \multicolumn{1}{c|}{13.4}    & \multicolumn{1}{c|}{17.5} & 9.0 ($33\%, 49\%$)  & \multicolumn{1}{c|}{8.5}        & 2.9 ($65\%$)  & \multicolumn{1}{c|}{10.7}               & 5.4 ($49\%$)  \\ \hline
\multicolumn{1}{|c|}{Deph $p=0.1$}  & \multicolumn{1}{c|}{23.7}    & \multicolumn{1}{c|}{29.9} & 21.0 ($11\%, 30\%$) & \multicolumn{1}{c|}{19.6}       & 15.2 ($22\%$) & \multicolumn{1}{c|}{15.0}               & 12.4 ($18\%$) \\ \hline
\multicolumn{1}{|c|}{Deph $p=0.2$}  & \multicolumn{1}{c|}{21.0}    & \multicolumn{1}{c|}{27.2} & 19.1 ($9\%, 30\%$)  & \multicolumn{1}{c|}{16.6}       & 10.6 ($36\%$) & \multicolumn{1}{c|}{13.5}               & 9.6 ($29\%$)  \\ \hline
\multicolumn{1}{|c|}{Deph $p=0.4$}  & \multicolumn{1}{c|}{15.2}    & \multicolumn{1}{c|}{21.4} & 14.8 ($3\%, 31\%$)  & \multicolumn{1}{c|}{9.6}        & 5.3 ($45\%$)  & \multicolumn{1}{c|}{10.1}               & 4.3 ($57\%$)  \\ \hline
\multicolumn{1}{|c|}{Damp $p=0.1$}  & \multicolumn{1}{c|}{25.4}    & \multicolumn{1}{c|}{31.1} & 20.8 ($18\%, 33\%$) & \multicolumn{1}{c|}{20.9}       & 15.0 ($28\%$) & \multicolumn{1}{c|}{13.0}               & 12.5 ($3\%$)  \\ \hline
\multicolumn{1}{|c|}{Damp $p=0.2$}  & \multicolumn{1}{c|}{24.4}    & \multicolumn{1}{c|}{29.7} & 18.7 ($23\%, 37\%$) & \multicolumn{1}{c|}{19.2}       & 11.1 ($42\%$) & \multicolumn{1}{c|}{9.0}                & 7.1 ($22\%$)  \\ \hline
\multicolumn{1}{|c|}{Damp $p=0.4$}  & \multicolumn{1}{c|}{22.1}    & \multicolumn{1}{c|}{26.2} & 13.9 ($37\%, 47\%$) & \multicolumn{1}{c|}{15.4}       & 5.6 ($64\%$)  & \multicolumn{1}{c|}{4.9}                & 2.9 ($40\%$)  \\ \hline
\end{tabular}
\caption{Comparison of negativity values for the noisy gates of a variational circuit in standard representations and obtained using the developed optimization algorithm.
In brackets the relative improvement after optimization is shown.}
\label{tab:variat}
\end{table*}

\section{Comparing impacts of gate merging and frame dimension increasing for random brick-wall circuits} \label{sec:random_circs}

\begin{figure}
    \centering
    \includegraphics[width=\linewidth]{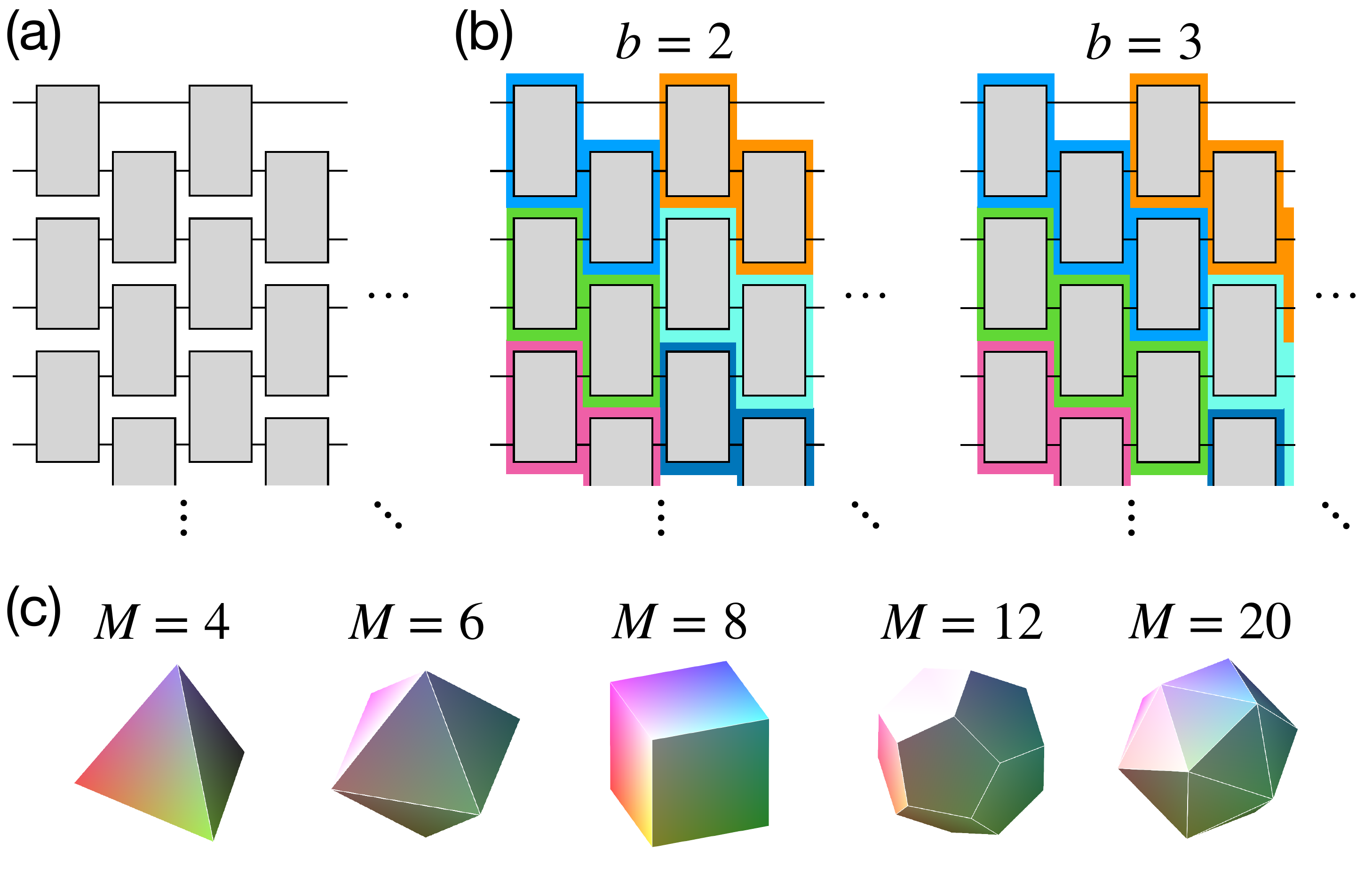}
    \caption{\BLUE{In (a) the brick-wall structure of the considered random circuit is shown. 
    Each two-qubit unitary is independently sampled according to the Haar measure.
    An application of each two-qubit unitary is accompanied by depolarizing noise of strength $p$ acting on affected qubits.
    In (b) the idea of gate merging for $b=2$ and $3$ is demonstrated.
    Two-qubit channels in blocks of the same color are multiplied together to obtain $(b+1)$-qubit channels.
    In (c) the list of regular polyhedrons used for constructing symmetric single-qubit frames $\eMap(M,R)$ is presented.}
    }
    \label{fig:gate_merging}
\end{figure}

In this section, we apply our approach to random circuits, which attracted significant attention in the framework of demonstrating quantum advantage~\cite{Martinis2019} and various problems in quantum many-body physics~\cite{fisher2023random}.
We consider a chain of qubits affected by a number of noisy two-qubit gates applied within a brick-wall pattern, as shown in Fig.~\ref{fig:gate_merging}(a).
The quantum channel $\Phi_{\hat U}(p)$ of each gate, which in an absence of noise realizes a unitary ${\hat U}$, is obtained as a composition of the corresponding ideal two-qubit unitary channel and depolarizing channel of strength $p$ acting on both qubits:
$\Phi_{\hat U}(p)=\Phi_{\hat U}\left(\{\hat A_i^{\rm depol}(p)\}_i\right)$
[see Eq.~\eqref{eq:two-qubit_dec}].
Unitary operators ${\hat U}$ of two-qubit gates are independently sampled from the Haar distribution.
In the case of leveraging the same generalized single-qubit frame $\eMap$ for all qubits, the total negativity of the circuit consisting of $N_{\rm gates}$ gates can be estimated as 
\begin{equation}
    \NegTot(\eMap,p) \approx N_{\rm gates} \cdot \langle \Neg\left(\eMap^{\otimes 2},\Phi_{\hat U}(p)\right) \rangle_{\rm Haar},
\end{equation}
where $\langle \cdot \rangle_{\rm Haar}$ denotes an averaging over Haar-random two-qubit unitaries $\hat{U}$.

Among with varying the form of the representation $\eMap$, which is considered in the previous sections, an alternative approach to decrease the negativity is joining channels together as shown in Fig.~\ref{fig:gate_merging}(b) and considering the negativity of composite blocks.
Along the lines of Ref.~\cite{Koukoulekidis_2022}, we refer to this approach as a gate merging.
This approach consists of composing gates together so that the negativity of the composite gate becomes smaller.
We note that in contrast to Ref.~\cite{Koukoulekidis_2022}, where special techniques of choosing a merging pattern are developed, here we consider a simplified straightforward version, which is motivated by the randomness of the gates.
After merging $b$ connected channels together, the total negativity can be estimated as
\begin{equation}
    \NegTot(\eMap, p, b) \approx \frac{N_{\rm gates}}{b} \cdot \langle \Neg\left(\eMap^{\otimes b+1},{\cal U}(p,b)\right) \rangle_{\rm Haar},
\end{equation}
where $\langle \cdot \rangle_{\rm Haar}$ stands for averaging over $(b+1)$-qubit channels of the following form:
\begin{equation}
    {\cal U}(p,b)=\left(\Id^{\otimes b-1} \otimes \Phi_{{\hat U_b}}(p)\right)\circ \ldots \circ \left(\Phi_{\hat U_1}(p)\otimes \Id^{\otimes b-1}\right),   
\end{equation}
which is obtained from $b$ independently distributed Haar-random two-qubit unitaries $\hat U_1,\ldots, \hat U_b$ (here, $\Id$ stands for the single-qubit identity channel).
According to Eq.~\eqref{eq:neg_of_join}, the negativity of a block can not exceed the sum of components' negativities, so $\NegTot(\eMap, p, b)$ is a non-increasing function of block size $b$.

The price for possible reduction of the negativity, is the exponential enlargement of the classical computational space per a single two-qubit unitary given by 
\begin{equation} \label{eq:size}
    {\sf size}(M,b) = \frac{1}{b} M^{2(b+1)}.
\end{equation}
Total space required for storing all quasi-stochastic matrices of the circuit in the memory of a classical computer is proportional to $N_{\rm gates}\times{\sf size}(M,b)$.
The expression of ${\sf size}(M,b)$ captures ``the curse of dimensionality'' and limits increasing of the parameter of $b$ for minimizing the negativity.

Next we return to the question of selecting the generalized frame $\eMap$.
Taking into account the uniform randomness of gates and isotropic behavior of the depolarizing channel, we consider a set of symmetric single-qubit frames $\eMap(M,R)$ whose synthesis operators are given by
\begin{equation}
    \eOp_i(M,R) = \frac{\Idop}{2} + R \sum_{j=1}^3 r_i^j(M)\hat\sigma_j,
\end{equation}
where $r_i^j(M)$ is a $j$-th component  of the $i$-th vertex of a three-dimensional $M$-vertex regular polyhedron inscribed in a unit sphere ($j=1,2,3$, $i=1,2,\ldots,M$) and $R>0$ is a parameter determining the Bloch vector radius of $\eOp_i(M,R)$.
We consider all the regular  three-dimensional polyhedrons, namely, tetrahedron ($M=4$), octahedron ($M=6$), cube ($M=8$), icosahedron ($M=12$), and dodecahedron ($M=20$), shown in Fig.~\ref{fig:gate_merging}(c).

Together with the expected reduction of the negativity as a result of the increasing of $M$, there is also a growth of the value of ${\sf size}(M,b)$, which determines the amount of classical memory required for storing a quasistochastic representation of the circuit.
We study the impact on the negativity decrease by varying the block size $b$ and dimension $M$ of symmetric frames $\eMap(M,R)$, given an upper limit on ${\sf size}(M,b)$.
In particular, we consider the value of normalized single-block negativity
\begin{equation} \label{eq:single-block-neg}
    \Neg_{\rm block}(M,b,p):=\frac{1}{b}
    \min_{R} \langle \Neg(\eMap(M,R)^{\otimes b+1},{\cal U}(p,b) ) \rangle_{\rm Haar},
\end{equation}
which determines the total negativity of the circuit, approximately given by $ N_{\rm gates}\Neg_{\rm block}(M,b,p)$.

\begin{figure}
    \centering
    \includegraphics[width=\linewidth]{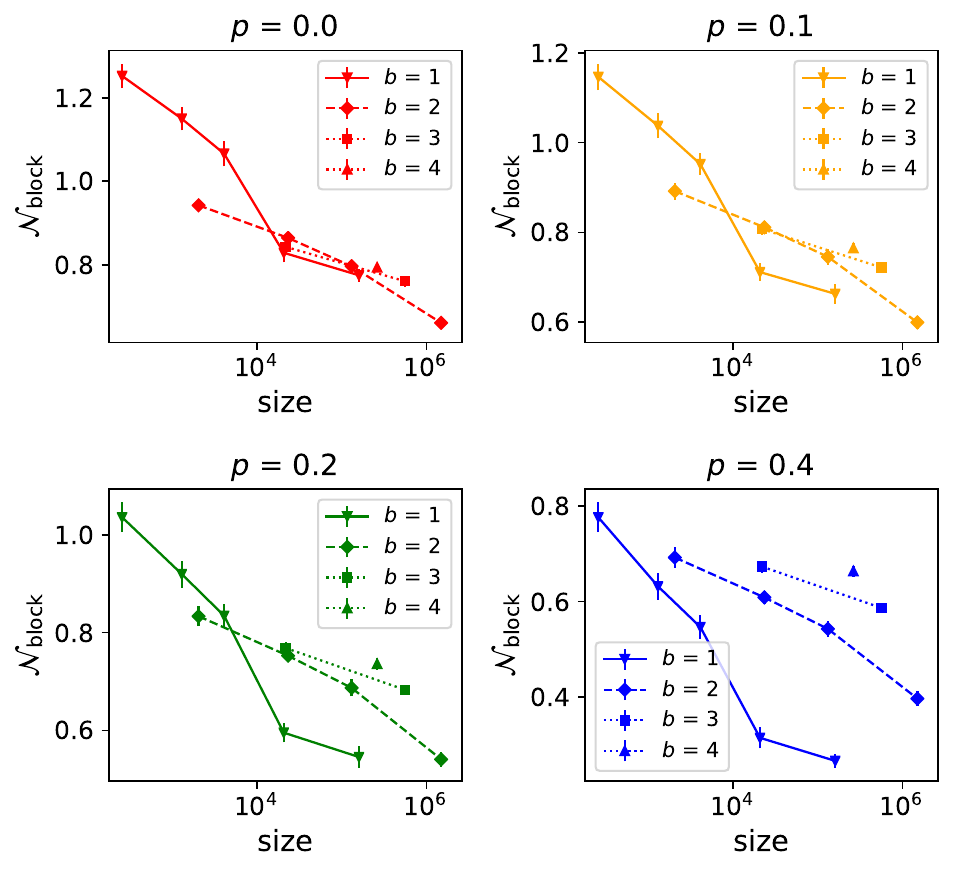}
    \caption{\BLUE{The normalized single-block negativity $\Neg_{\rm block}(M,b,p)$ as functiton of ${\sf size}(M,b)$ for different configurations of regular frame dimension $M$, block size $b$, and strength of depolarization $p$. 
    The limitation on the normailized number of elements in the resulting quasi-stochastic matrices ${\sf size}(M,b)<3 \cdot 10^6$ is considered.
    Vertical lines denote a standard deviation among the processed data.}}
    \label{fig:avg_merge_negs}
\end{figure}

In Fig.~\ref{fig:avg_merge_negs} we demonstrated the behavior of $\Neg_{\rm block}(M,b,p)$, which is obtained by processing a number of Haar-random two-qubit unitaries affected by depolarization of strengths $p\in\{0,0.1,0.2,0.4\}$, for values of $M$ and $b$ satisfying  ${\sf size}(M,b)<2 \cdot 10^6$.
For $b=1$, all values of $M\in\{4,6,8,12,20\}$ are considered, while for $b=2$, $3$, and $4$, $M\in\{4,6,8,12\}$, $M\in\{4,6\}$, and $M=4$ are taken correspondingly.
Notably, optimal values of the radius parameter $R$ typically  appear in the region $[0.85, 1.1]$ (for more details, see Appendix~\ref{appendix:radius_opt}).
The results of Fig.~\ref{fig:avg_merge_negs} indicate that depolarization affects the relation between impacts of gate merging and increasing of $M$.
In the case of the absence of decoherence ($p=0$), we see a competitive behavior: The configuration $b=2$, $M=4$ outperforms $b=1$, $M=8$, yet $b=1$, $M=12,20$ are very close to  $b=2$, $M=6,8$ and $b=3$, $M=4,6$.
However, with as increase of $p$, we observe a repelling of lines corresponding to different $b$, and for $p=0.4$ the impact of increasing the dimensionality clearly outperforms the gate merging.

To summarize, we observe that the impact on negativity of taking a generalized frame of higher dimensionality can be stronger compared to the gate merging, given a limit on the resulting size of the circuit representation in a memory of a classical computer.
Within the considered setup, the advantage of frame dimension increasing grows with increasing the noise.
We leave the question about influence of a decoherence of the more general form on the relation between the frame dimension increasing the gate merging for the future research.

\section{Conclusion and outlook}\label{sec:conclusion}

Here we summarize the main results of our paper.
We have presented a generalization to the theory of quasiprobability representations, which allows for the study of the optimal quasiprobability descriptions of quantum circuits in the overcomplete case.
We have developed an algorithm for the minimization of negativity for arbitrary quantum circuits.
We have tested a performance of the algorithm in minimizing negativity of qubit Clifford+T blocks, and layers of variational quantum circuits.
We have shown how negativity decreases with the growth of representation dimensionality $M$ and increasing of decoherence strength.
We have observed a rather significant drop of negativity when transferring from $M=4$ up to the $M=8$ case.
Notably, in all cases except ideal Clifford+T blocks, the obtained optimized frames are different from stabilizer states-based frame ($M=6$), and $\Lambda$ polytope $(M=8)$.
Moreover, we have shown that the negativity of a circuit can be minimized by combining increasing frame dimension and gate merging technique.
We obtained that in the case of depolarized random brick-wall circuits, increasing dimension of symmetric frames appears to more efficient compared to the gate merging in the case of a strong depolarizing noise (given a restriction on resulting size of quasistochastic matrices).

As we have seen, the further increase of $M$ does not improve negativity in the case of Clifford+T blocks, however, but has a little impact for a variational circuit. 
These numerical experiments clearly indicate that the negativity of arbitrary circuits asymptotically tends to some nonzero limit with increasing $M$.
It is important to note that the minimization has been performed over factorized frames with identical analysis operators for each qubit.
We believe the developed approach can be used for deeper study of the origin of the potential computational advantage brought by NISQ devices.

We also suppose that the theory of asymptotic geometric analysis can be applied to study the properties of asymptotic behavior of negativity.
As we have seen, when increasing frame dimension $M$, the quasiprobability representation can become ``more symmetric''. 
In the theory of frames this is expressed formally with the concept of complex projective designs~\cite{Casazza_2012, Dankert2005, Klappenecker2005}. 
It may be beneficial to develop similar concepts for generalized quasiprobability representations, e.g., we suspect the notion of frame potential is related to an average negativity of Haar-random unitaries.

\section*{Acknowledgments}
We thank I.A. Luchnikov and J.A. López-Saldívar for useful comments.
The work was supported by Russian Science Foundation Grant No. 19-71-10091 (development of generalized frames). 
The work is also supported by the Priority 2030 program at NIST ``MISIS'' under project K1-2022-027.

\section*{Data and code availability}

The source code of the developed optimization procedure as well as obtained frames are available at~\cite{supplementary_new}.

\appendix

\section{Examples of frame-based quasiprobability representations} \label{appendix:frame-based}

\subsection{SIC-POVM-based representation} \label{appendix:SIC}

One can define a representation based on SIC-POVMs~\cite{Renes_2004, Zhu_2016, Kiktenko_2020}.
Let us consider a two-dimensional Hilbert space with Pauli matrices $\{\hat{\mathbb{1}},\hat\sigma_1,\hat\sigma_2,\hat\sigma_3\}$. 
The \emph{SIC-POVM representation} is a MIC ($M = D = 4$) representation $(\EMap,\eMap)$ given by the operators
\begin{equation}
\begin{aligned}
	\EOp_1 = \frac{1}{4}\left(\hat{\mathbb{1}} - \frac{1}{\sqrt{3}}\hat\sigma_1 + \frac{1}{\sqrt{3}}\hat\sigma_2 + \frac{1}{\sqrt{3}}\hat\sigma_3\right), \\
	\EOp_2 = \frac{1}{4}\left(\hat{\mathbb{1}} + \frac{1}{\sqrt{3}}\hat\sigma_1 - \frac{1}{\sqrt{3}}\hat\sigma_2 + \frac{1}{\sqrt{3}}\hat\sigma_3\right), \\
	\EOp_3 = \frac{1}{4}\left(\hat{\mathbb{1}} + \frac{1}{\sqrt{3}}\hat\sigma_1 + \frac{1}{\sqrt{3}}\hat\sigma_2 - \frac{1}{\sqrt{3}}\hat\sigma_3\right), \\
	\EOp_4 = \frac{1}{4}\left(\hat{\mathbb{1}} - \frac{1}{\sqrt{3}}\hat\sigma_1 - \frac{1}{\sqrt{3}}\hat\sigma_2 - \frac{1}{\sqrt{3}}\hat\sigma_3\right),
\end{aligned}
\end{equation}
and
\begin{equation}
\begin{aligned}
	\eOp_1 = \frac{1}{2}\left(\hat{\mathbb{1}} - \sqrt{3}\hat\sigma_1 + \sqrt{3}\hat\sigma_2 + \sqrt{3}\hat\sigma_3\right), \\
	\eOp_2 = \frac{1}{2}\left(\hat{\mathbb{1}} + \sqrt{3}\hat\sigma_1 - \sqrt{3}\hat\sigma_2 + \sqrt{3}\hat\sigma_3\right), \\
	\eOp_3 = \frac{1}{2}\left(\hat{\mathbb{1}} + \sqrt{3}\hat\sigma_1 + \sqrt{3}\hat\sigma_2 - \sqrt{3}\hat\sigma_3\right), \\
	\eOp_4 = \frac{1}{2}\left(\hat{\mathbb{1}} - \sqrt{3}\hat\sigma_1 - \sqrt{3}\hat\sigma_2 - \sqrt{3}\hat\sigma_3\right).
\end{aligned}
\end{equation}
This representations has a lot of symmetry and quite often is used in numerical simulations of quantum systems~\cite{Luchnikov_2019, Koukoulekidis_2022}.

\subsection{Wigner representation} \label{appendix:Wigner}

Probably the most important example of finite-dimensional quasiprobability representation comes with the use of the Wigner function~\cite{Gibbons_2004, Gross_2006}. 
This representation gives a convenient statistical model for simulating Clifford circuits over qudits~\cite{Veitch_2012}.

Suppose $d$ is an odd prime dimension, and let $\Hil_d$ be a Hilbert space representing a qudit with computational basis $\{\ket{0},\dots,\ket{d-1}\}$. 
Let $\omega = \exp(i 2\pi / d)$ be the $d$-th primitive root of unity.
The shift $\hat X$ and clock $\hat Z$ operators are defined as
\begin{equation}
	\begin{aligned}
	&\hat X\ket{k} = \ket{k+1 \mod d}, \\
	&\hat Z\ket{k} = \omega^k \ket{k}.
	\end{aligned}
\end{equation}
These operators commute as $\hat Z \hat X = \omega \hat X \hat Z$. 
The operators $\hat X$ and $\hat Z$ generate a Heisenberg-Weyl group if we include phases. 
A general element of the group is defined up to a phase as
\begin{equation}
	\hat T_{(u_Z,u_X)} = \omega^{- u_Z u_X / 2} \hat Z^{u_Z} \hat X^{u_X},
\end{equation}
where the vector $(u_X, u_Z)$ lies in a phase space $\mathbb{Z}_d\times\mathbb{Z}_d$. 
The phase point operators are given by the symplectic Fourier transform of Heisenberg-Weyl operators:
\begin{equation}
	\hat A_0 = \frac{1}{d}\sum_{u} \hat T_u, \quad \hat A_u = \hat T_u \hat A_0 \hat T_u^\dag,
\end{equation}
where $u \in \mathbb{Z}_d\times\mathbb{Z}_d$ is a point in the phase space. 
The operators $\{A_u\}_u$ are informationally complete and satisfy $\Tr(A_u A_v) = d \delta_{u,v}$. 
Therefore, they constitute a MIC quasiprobability representation, if we define for all $u \in \mathbb{Z}_d\times\mathbb{Z}_d$:
\begin{equation}
	\hat e_u = \hat A_u, \quad \hat E_u = \frac{1}{d} \hat A_u.
\end{equation}
This representation is of great interest due to the fact that Clifford gates are represented as permutation matrices, which makes them especially easy to simulate. 
The properties of negativity in this representation are also deeply connected with the abundance of magic in the circuit~\cite{Veitch_2012, Veitch_2014}.
In fact, this representation is in a certain way unique~\cite{Schmid_2022}, and no such representation exists if $d$ is even.
Nevertheless, recently there were found some schemes for the simulation of qubit circuits in a similar way~\cite{Zurel_2020,Zurel_2021,Raussendorf_2020} (see Appendix~\ref{appendix:Lambda}).

\subsection{Wootters representation} \label{appendix:Wootters}

Wootters in Ref.~\cite{Wootters_1987} defined a phase space representation for qubits. 
The \emph{Wootters representation} is a MIC ($M = D = 4$) representation $(\EMap,\eMap)$ given by the operators
\begin{equation}
\begin{aligned}
  \eOp_0 = \frac{1}{2}\left(\hat{\mathbb{1}} + \hat\sigma_1 + \hat\sigma_2 + \hat\sigma_3\right), \\
  \eOp_1 = \frac{1}{2}\left(\hat{\mathbb{1}} + \hat\sigma_1 - \hat\sigma_2 - \hat\sigma_3\right), \\
  \eOp_2 = \frac{1}{2}\left(\hat{\mathbb{1}} - \hat\sigma_1 + \hat\sigma_2 - \hat\sigma_3\right), \\
  \eOp_3 = \frac{1}{2}\left(\hat{\mathbb{1}} - \hat\sigma_1 - \hat\sigma_2 + \hat\sigma_3\right).
\end{aligned}
\end{equation}
and
\begin{equation}
\begin{aligned}
  \EOp_0 = \frac{1}{4}\left(\hat{\mathbb{1}} + \hat\sigma_1 + \hat\sigma_2 + \hat\sigma_3\right), \\
  \EOp_1 = \frac{1}{4}\left(\hat{\mathbb{1}} + \hat\sigma_1 - \hat\sigma_2 - \hat\sigma_3\right), \\
  \EOp_2 = \frac{1}{4}\left(\hat{\mathbb{1}} - \hat\sigma_1 + \hat\sigma_2 - \hat\sigma_3\right), \\
  \EOp_3 = \frac{1}{4}\left(\hat{\mathbb{1}} - \hat\sigma_1 - \hat\sigma_2 + \hat\sigma_3\right).
\end{aligned}
\end{equation}
This representation is a qubit analog of the discrete Wigner function. 
We note, however, that it does not have the full strength of the Wigner representation for qudit systems.

\section{Examples of generalized quasiprobablity representations} \label{appendix:generalized}

\subsection{Stabilizer states-based representation} \label{appendix:stabilizer}

One may consider a representation $\eMap$ defined by a set of all stabilizer states on $n$ qubits.
For a single qubit this representation is made of
\begin{equation}
\begin{aligned}
	&\eOp_1 = \ketbra{+}{+},    && \hat e_2 = \ketbra{-}{-}, \\
	&\eOp_3 = \ketbra{+\imath}{+\imath},  && \hat e_4 = \ketbra{-\imath}{-\imath}, \\
	&\eOp_5 = \ketbra{0}{0},    && \hat e_6 = \ketbra{1}{1},
\end{aligned}
\end{equation}
where $\ket{\pm}\Def 2^{-1/2}(\ket{0}\pm\ket{1})$ and $\ket{\pm\imath}\Def 2^{-1/2}(\ket{0}\pm\imath\ket{1})$.
For $n$ qubits, these representations are huge \cite{Aaronson_2004, Gross_2006}:
\begin{equation}
	M = 2^n\prod_{k=1}^{n}(2^k+1) = 2^{\frac{1}{2}n^2+\frac{3}{2}n+O(1)}.
\end{equation}
Nevertheless, there are a number of ways it can be used to simulate quantum circuits~\cite{Howard_2017, Seddon_2021}. 
The negativity of this representation is usually called \emph{robustness of magic}~\cite{Howard_2017, Heinrich_2019}.

\subsection{\texorpdfstring{$\Lambda$}{Lambda} polytope representation} \label{appendix:Lambda}

Recently, there were found noncontextual hidden variable models describing quantum computation with magic states~\cite{Zurel_2020,Okay_2021,Zurel_2021,Raussendorf_2020}. 
These models use the notion of \emph{$\Lambda$ polytope} on $n$ qubits defined as follows:
\begin{equation}
	\Lambda_n \Def \{\rho \in \THil : \bra{\sigma}\rho\ket{\sigma}\geq 0 \text{ for all stabilizer }\ket{\sigma} \}
\end{equation}
It may be thought of as a \emph{polar set} to a set of all stabilizer projections.
The polytope $\Lambda_n$ is a set of \emph{quasistates} that behaves ``well'' under Clifford unitaries and Pauli measurements. 
At the same time, the polytope $\Lambda_n$ contains all the states on $n$ qubits, including magic states.

We can consider the representation $\eMap$ made of all the extreme points of this polytope. In this representation all the states, Clifford unitaries, and Pauli measurements are positive, 
so the Clifford+magic circuits have zero negativity. 
At the same time, the structure of this representation for arbitrary $n$ is very complicated, and the number $M$ is superexponential~\cite{Zurel_2023}.

For a single qubit $n=1$ this representation contains $M=8$ quasistates:
\begin{equation}
\begin{aligned}
  \eOp_0 = \frac{1}{2}\left(\hat{\mathbb{1}} + \hat\sigma_1 + \hat\sigma_2 + \hat\sigma_3\right), \\
  \eOp_1 = \frac{1}{2}\left(\hat{\mathbb{1}} + \hat\sigma_1 + \hat\sigma_2 - \hat\sigma_3\right), \\
  \eOp_2 = \frac{1}{2}\left(\hat{\mathbb{1}} + \hat\sigma_1 - \hat\sigma_2 + \hat\sigma_3\right), \\
  \eOp_3 = \frac{1}{2}\left(\hat{\mathbb{1}} + \hat\sigma_1 - \hat\sigma_2 - \hat\sigma_3\right), \\
  \eOp_4 = \frac{1}{2}\left(\hat{\mathbb{1}} - \hat\sigma_1 + \hat\sigma_2 + \hat\sigma_3\right), \\
  \eOp_5 = \frac{1}{2}\left(\hat{\mathbb{1}} - \hat\sigma_1 + \hat\sigma_2 - \hat\sigma_3\right), \\
  \eOp_6 = \frac{1}{2}\left(\hat{\mathbb{1}} - \hat\sigma_1 - \hat\sigma_2 + \hat\sigma_3\right), \\
  \eOp_7 = \frac{1}{2}\left(\hat{\mathbb{1}} - \hat\sigma_1 - \hat\sigma_2 - \hat\sigma_3\right),
\end{aligned}
\end{equation}
which are the vertices of a cube circumscribed around the Bloch sphere.

\section{Optimization of radius parameter of symmetric single-qubit frames} \label{appendix:radius_opt}

\begin{figure}
    \centering
    \includegraphics[width=\linewidth]{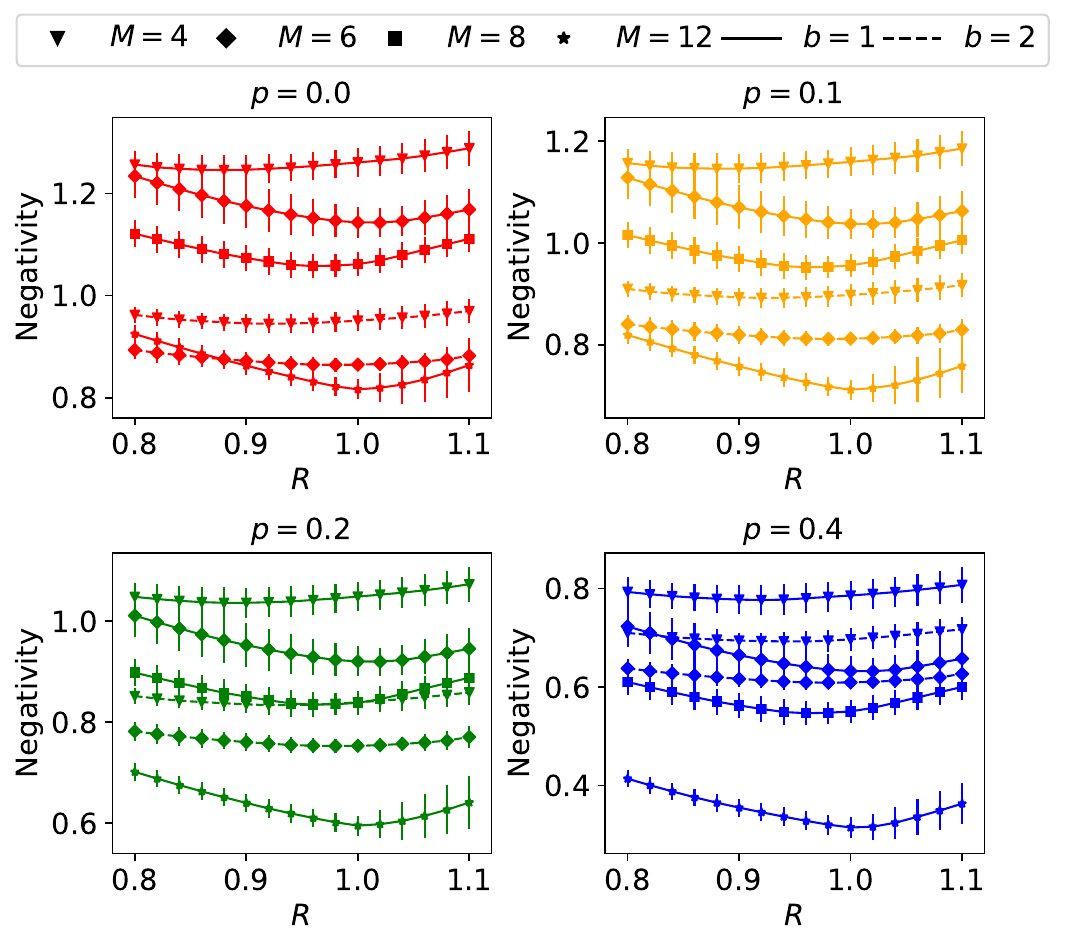}
    \caption{Dependence of $\Neg(\eMap(M,R)^{\otimes b+1},{\cal U}(p,b))\rangle_{\rm Haar}$ on $R$ for various configurations of $M$, $b$, and $p$.
    Vertical lines denote a standard deviation among the processed data.}
    \label{fig:R-dep}
\end{figure}

Here we provide some additional details related to optimization over parameter $R$ of the normalized single-block negativity~\eqref{eq:single-block-neg} studied in Sec.~\ref{sec:random_circs}.
In Fig.~\ref{fig:R-dep} we show a dependence of $\Neg(\eMap(M,R)^{\otimes b+1},{\cal U}(p,b) ) \rangle_{\rm Haar}$ on $R$ for various configurations of $M$, $b$, and $p$.
We see that positions of extreme are actually determined by values of $b$ and $M$ for all considered decoherence strengths $p$, and are spread inside the region $[0.85, 1.1]$.
Due to a smooth behavior of the curves, $R=0.95$ can be used as a rather good approximation of the optimal value of the radius.

\bibliographystyle{apsrev4-1}
\bibliography{bibliography}

\end{document}